\documentclass{ws-rv9x6}
\usepackage{subfigure}   
\usepackage{ws-rv-thm}   
\usepackage{ws-rv-van}   
\usepackage[colorlinks=true,urlcolor=blue]{hyperref}

\makeindex

\begin{document}

\setcounter{chapter}{4}
\chapter[Spin-orbit-coupled Bose--Einstein
condensates]{Spin-orbit-coupled Bose--Einstein
condensates}\label{ra_ch1}

\author[Y. Li and G. I. Martone and S. Stringari]{Yun Li$^{1,2}$
and Giovanni I. Martone$^1$ and Sandro Stringari$^1$}
\address{$^1$\hspace*{-1pt}Dipartimento di Fisica, Universit\`{a}
di Trento and INO-CNR BEC Center, I-38123 Povo, Italy\\
$^2$\hspace*{-1pt}Centre for Quantum Technologies, National University of
Singapore,\\ 3 Science Drive 2, 117542, Singapore}

\begin{abstract}
The recent realization of synthetic spin-orbit coupling represents an
outstanding achievement in the physics of ultracold quantum gases.
In this review we explore the properties of a spin-orbit-coupled
Bose--Einstein condensate with equal Rashba and Dresselhaus strengths.
This system presents a rich phase diagram, which exhibits a
tricritical point separating a zero-momentum phase, a spin-polarized
plane-wave phase, and a stripe phase. In the stripe phase
translational invariance is spontaneously broken, in analogy with
supersolids. Spin-orbit coupling also strongly affects the dynamics
of the system. In particular, the excitation spectrum exhibits
intriguing features, including the suppression of the sound velocity,
the emergence of a roton minimum in the plane-wave phase, and the
appearance of a double gapless band structure in the stripe phase.
Finally, we discuss a combined procedure to make the stripes visible
and stable, thus allowing for a direct experimental detection.
\end{abstract}

\body

\section{Introduction}
\label{sec:introduction}

A large variety of exotic phenomena in solid-state systems
can take place when their constituent electrons are coupled
to an external gauge field, or in the presence of strong
spin-orbit coupling. For example, magnetic fields influencing
the motion of the electrons are at the base of the well-known
quantum Hall effect \cite{Klitzing1986}, whereas
spin-orbit coupling, i.e., the coupling between an electron's
spin and its momentum, is crucial for topological
insulators \cite{Hasan2010,Qi2011}, Majorana fermions
\cite{Wilczek2009}, spintronic devices \cite{Koralek2009}, etc.
Ultracold
atomic gases are good candidates to investigate
these interesting quantum phenomena. In this respect, the
main difficulty arises from the fact that atoms are neutral
particles, and consequently they cannot be coupled to a
gauge field. In addition, they do not exhibit any coupling
between their spin and their center-of-mass motion.

In the last few years there have been several proposals to
realize artificial gauge fields for quantum gases, thus
overcoming the problem of their neutrality \cite{Dalibard2011}.
One of these schemes relies on the notion of geometric phase
\cite{Berry1984}, which emerges when the motion of a particle
with some internal level structure is slow enough, so that
the particle follows adiabatically one of these levels.
In such conditions, the particle experiences an effective
vector potential. In ultracold atomic gases, several
methods to implement these ideas exploit the space-dependent
coupling of the atoms with a properly designed configuration
of laser beams; the synthetic gauge field arises when the
system follows adiabatically one of the local eigenstates of
the light-atom interaction Hamiltonian (dressed states)
\cite{Ruseckas2005,Zhu2006,Gunter2009,Cooper2010}. Other
approaches are also possible, such as the periodic shaking
of an optical lattice with special frequencies, which couples
different Bloch bands \cite{Hauke2012}.

Since 2009, several experiments have been successful in realizing
ultracold atomic gases coupled to artificial gauge fields
\cite{Spielman2009,Lin2009PRL,Lin2009Nature,Struck2011,Parker2013}.
For instance, in the experiment of Ref.~\citen{Lin2009PRL} a
space-dependent atom-light coupling was employed to simulate an
effective magnetic field exerting a Lorentz-like force on neutral
bosons; this procedure has been used to generate quantized
vortices in Bose--Einstein condensates (BECs).

Another interesting situation occurs when the local dressed states
are degenerate, giving rise to spin-orbit-coupled configurations. In
particular, by using a suitable arrangement of Raman lasers, the
authors of Ref.~\citen{Lin2011} managed to engineer a one-dimensional
spin-orbit coupling, characterized by equal Rashba \cite{Bychkov1984}
and Dresselhaus \cite{Dresselhaus1955} strengths, on a neutral
atomic BEC. The same scheme has been subsequently extended
to realize spin-orbit-coupled Fermi gases \cite{Wang2012,Cheuk2012}.

These first experimental achievements have stimulated a growing
interest in this field of research, resulting in a wide number of
papers devoted to artificial gauge fields and, more specifically,
to spin-orbit-coupled quantum gases, both from the theoretical
and the experimental side. In this review we will focus on the
properties of Bose--Einstein condensates with the kind of
spin-orbit coupling first realized by the NIST team \cite{Lin2011}.
Readers who are interested in a broader overview
about spin-orbit-coupled quantum gases and, more generally, about
artificial gauge fields on neutral atoms, can refer to some recent
reviews \cite{Dalibard2011,Galitski2013,Goldman2014,Zhai2014} and
references therein.

This paper is organized as follows. In Sec.~\ref{sec:Ground_state}
we illustrate the quantum phase diagram of the system. The dynamic
behavior of the gas in the two uniform phases is studied in
Sec.~\ref{sec:Dynamic_uniform}. Section~\ref{sec:Collect_trap} deals
with the collective excitations in the presence of harmonic trapping.
Section~\ref{sec:Excitation_stripe} is entirely devoted to the phase
exhibiting density modulations in the form of stripes: we discuss
both the ground state and the excitation spectrum, and we illustrate
a procedure allowing for the direct observation of the stripes.
Finally, in Sec.~\ref{sec:Conclusion} we report some brief
concluding remarks.

\section{Ground-state Phase Diagram}
\label{sec:Ground_state}

\subsection{Single-particle picture}
\label{subsec:Single_particle}
\begin{figure}[t]
\centering
\includegraphics[scale=1]{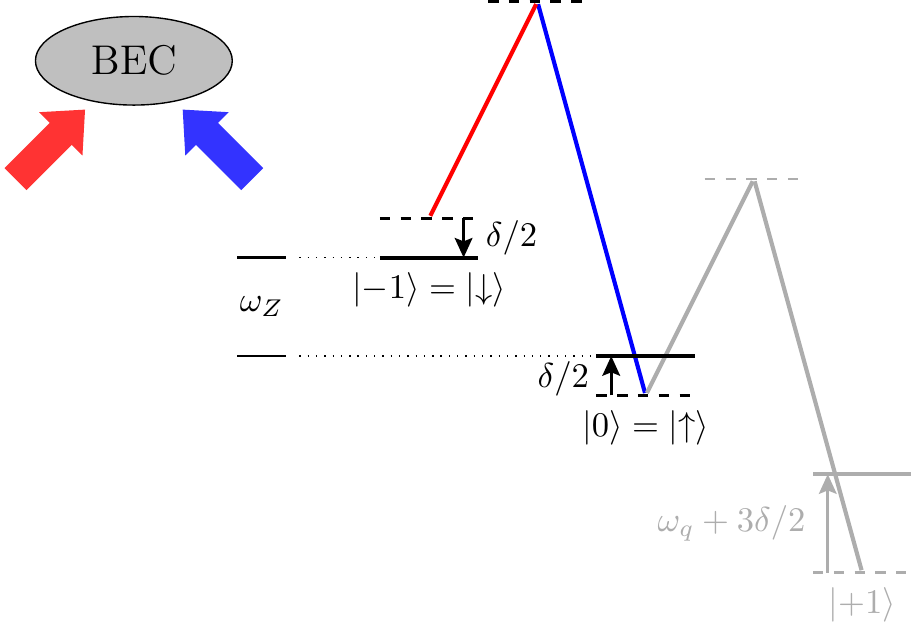}
\caption{Level diagram. Two Raman lasers with orthogonal linear
polarizations couple the two states $\left|\uparrow\right\rangle
=\left|m_F=0\right\rangle$ and $\left|\downarrow\right\rangle=
\left|m_F=-1\right\rangle$ of the $F=1$ hyperfine manifold of
$^{87}$Rb, which differ in energy by a Zeeman splitting $\omega_Z$.
The lasers have frequency difference $\Delta\omega_L = \omega_Z +
\delta$, where $\delta$ is a small detuning from the Raman
resonance. The state $\left|m_F=1\right\rangle$ can be neglected
since it has a much larger detuning, due to the quadratic Zeeman
shift $\omega_q$.} \label{fig:level_diagram}
\end{figure}

The experimental setup employed in Ref.~\citen{Lin2011} to realize
spin-orbit coupling consists of a $^{87}$Rb Bose--Einstein condensate
in the $F=1$ hyperfine manifold, with a bias magnetic field
providing a nonlinear Zeeman splitting between the three levels of the manifold.
The BEC is coupled to the field of two Raman lasers having orthogonal
linear polarizations, frequencies $\omega_L$ and
$\omega_L + \Delta\omega_L$, and wave vector difference ${\bf k}_0 =
k_0 \hat{\bf e}_x$, with $\hat{\bf e}_x$ the unit vector along the $x$
direction. The laser field induces transitions between the three states
characterized by a Rabi frequency $\Omega$ fixed by the intensity
of the lasers. This Raman process is illustrated schematically
in Fig.~\ref{fig:level_diagram}. The frequency splitting $\omega_Z$
between the states $\left|F=1,m_F=0\right\rangle$ and
$\left|F=1,m_F=-1\right\rangle$ is chosen to be very close to the
frequency difference $\Delta\omega_L$ between the two lasers,
while the separation $\omega_Z-\omega_q$ between
$\left|F=1,m_F=0\right\rangle$ and $\left|F=1,m_F=1\right\rangle$
contains a large additional shift from Raman resonance due to the
quadratic Zeeman effect. This implies that the state
$\left|m_F=1\right\rangle$ can be neglected, and we are left with
an effective spin-$1/2$ system, with the two spin states given by
$\left|\uparrow\right\rangle = \left|m_F=0\right\rangle$ and
$\left|\downarrow\right\rangle = \left|m_F=-1\right\rangle$.
The single-particle Hamiltonian of this system takes the form
(we set $\hbar = m = 1$)
\begin{equation}
h_0 = \frac{{\bf p}^2}{2} +\frac{\Omega}{2}\sigma_x \cos(2k_0 x -
\Delta\omega_L t)+\frac{\Omega}{2} \sigma_y \sin(2 k_0 x -
\Delta\omega_L t) - \frac{\omega_Z}{2} \sigma_z ,
\label{eq:h_0}
\end{equation}
where $\sigma_k$ with $k=x,\,y,\,z$ denotes the usual $2\times 2$
Pauli matrices. Hamiltonian (\ref{eq:h_0}) is not translationally
invariant but exhibits a screwlike symmetry, being invariant with
respect to helicoidal translations of the form
$e^{id(p_x-k_0\sigma_z)}$, consisting of a combination of a rigid
translation by distance $d$ and a spin rotation by angle $-d k_0$
around the $z$ axis.

Let us now apply the unitary transformation $e^{i\Theta \sigma_z/2}$,
corresponding to a position and time-dependent rotation in spin space
by the angle $\Theta =2 k_0 x - \Delta \omega_L t$, to the wave
function obeying the Schr\"{o}dinger equation. As a consequence of the
transformation, the single-particle Hamiltonian (\ref{eq:h_0}) is
transformed into the translationally invariant and time-independent
form
\begin{equation}
h_0^{\rm SO}= \frac{1}{2}\left[\left(p_x-k_0 \sigma_z\right)^2 +
p_\perp^2 \right] + \frac{\Omega}{2} \sigma_x + \frac{\delta}{2}
\sigma_z .
\label{eq:h_0_SO}
\end{equation}
The spin-orbit nature acquired by the Hamiltonian results from the
noncommutation of the kinetic energy and the position-dependent
rotation, while the renormalization of the effective magnetic field
$\delta = \Delta\omega_L-\omega_Z$ results from the additional
time dependence exhibited by the wave function in the rotating
frame. The new Hamiltonian is characterized by equal contributions
of Rashba \cite{Bychkov1984} and Dresselhaus \cite{Dresselhaus1955}
couplings. It has the peculiar property of violating both parity and
time-reversal symmetry.
It is worth pointing out that the operator ${\bf p}$
entering Eq.~(\ref{eq:h_0_SO}) is the canonical momentum $-i\nabla$, with
the physical velocity being given by ${\bf v}_\pm= {\bf p} \mp k_0
\hat{\bf e}_x$ for the spin-up and spin-down particles. In
terms of ${\bf p}$ the eigenvalues of Hamiltonian (\ref{eq:h_0_SO})
are given by
\begin{equation}
\varepsilon_{\pm}({\bf p}) = \frac{p_x^2 + p_\perp^2}{2} + E_r \pm
\sqrt{\left(k_0 p_x - \frac{\delta}{2}\right)^2 + \frac{\Omega^2 }{4}} ,
\label{eq:E_single}
\end{equation}
where $E_r = k_0^2/2$ is the recoil energy. The
double-branch structure exhibited by the dispersion (\ref{eq:E_single})
reflects the spinor nature of the system.

\begin{figure}[t]
\centering
\includegraphics[scale=1]{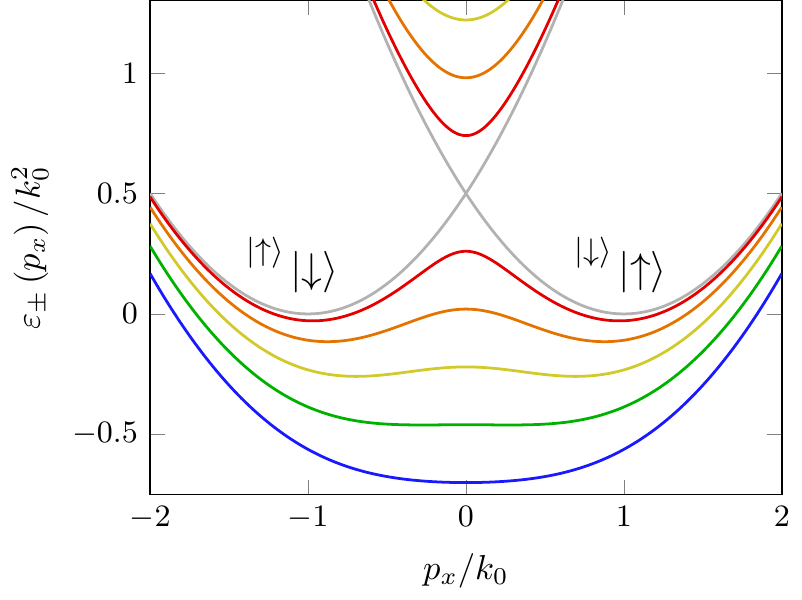}
\caption{Single-particle dispersion (\ref{eq:E_single}) at
$\delta=0$. Eigenenergies calculated for Raman coupling ranging
from $\Omega=0$ (grey) to $\Omega = 2.4 \, k_0^2$ (blue).
The two minima in the lower branch disappear at
$\Omega = 2k_0^2$.}
\label{fig:single_particle}
\end{figure}

We now focus on the case $\delta=0$ and $\Omega \geq 0$. In
Fig.~\ref{fig:single_particle} we plot the dispersion (\ref{eq:E_single})
as a function of $p_x$, for different values of $\Omega$. The lower
branch $\varepsilon_-({\bf p})$ exhibits, for $\Omega < 2k_0^2$, two
degenerate minima at momenta ${\bf p}= \pm k_1 \hat{\bf e}_x$ with
$k_1 = k_0 \sqrt{1 - \Omega^2/4k_0^4}$, both capable to host
Bose--Einstein condensation.
At larger values of $\Omega$ the spectrum has instead a single
minimum at ${\bf p}=0$. The effective mass of particles moving along
$x$, fixed by the relation $m/m^\ast = {\rm d}^2\varepsilon / {\rm d}
p_x^2$, also shows a nontrivial $\Omega$ dependence. Near the minimum
one finds
\begin{align}
\dfrac{m\;}{m^\ast} &= 1-\left(\frac{\Omega}{2k_0^2} \right)^2
&\hspace*{-2cm}\text{for} \quad \Omega < 2k_0^2 ,
\label{eq:mass_12}\\
\dfrac{m\;}{m^\ast} &= 1-\frac{2k_0^2}{\Omega}
&\hspace*{-2cm}\text{for} \quad \Omega > 2k_0^2 .
\label{eq:mass_3}
\end{align}
Thus, the effective mass exhibits a divergent behavior at $\Omega=2k_0^2$,
where the double-well structure disappears and the spectrum has a
$p_x^4$ dispersion near the minimum.

Before concluding the present section,
it is worth mentioning that a single-particle dispersion similar to
(\ref{eq:E_single}) can also be achieved by trapping the atoms in a
shaken optical lattice, as recently realized experimentally
\cite{Parker2013}. In such systems, different Bloch bands coupled
through lattice shaking bear several analogies with the spin states
involved in the Raman process described above \cite{Zheng2014}.

\subsection{Many-body ground state}
\label{subsec:Many_body}
We shall now illustrate how the peculiar features of the single-particle
dispersion (\ref{eq:E_single}) are at the origin of new interesting phases
in the many-body ground state of the BEC. For a gas of $N$ particles enclosed
in a volume $V$, in the presence of two-body interactions, the many-body
Hamiltonian takes the form
\begin{equation}
H = \sum_j h_0^{\rm SO}(j) + \sum_{\sigma,\,\sigma'} \frac{1}{2} \int
{\rm d}{\bf r} \, g_{\sigma\sigma'}\, \rho_{\sigma} ({\bf r})
\rho_{\sigma'} ({\bf r}),
\label{eq:H_N_body}
\end{equation}
where $h_0^{\rm SO}$ is given by (\ref{eq:h_0_SO}), $j=1,\ldots,N$ is
the particle index, and $\sigma,\,\sigma'$ are the spin indices
($\uparrow,\downarrow \,=\pm$) characterizing the two spin states.
The spin-up and spin-down density operators
entering Eq.~(\ref{eq:H_N_body}) are defined by
$\rho_\pm({\bf r}) = (1/2) \sum_j \left(1 \pm
\sigma_{z,j} \right) \delta({\bf r} - {\bf r}_j)$, while
$g_{\sigma\sigma'} = 4\pi a_{\sigma\sigma'}$ are the relevant coupling
constants in the different spin channels, with $a_{\sigma\sigma'}$
the corresponding $s$-wave scattering lengths.

To investigate the ground state of the system we resort to the
Gross--Pitaevskii mean-field approach, and we write the energy
functional associated to Hamiltonian (\ref{eq:H_N_body}) as
\begin{equation}
\begin{aligned}
E = &\int {\rm d}{\bf r} \, \Psi^\dagger({\bf r}) h_0^{\rm SO} \Psi({\bf r})
\\
&{} +\int {\rm d}{\bf r}\left[ \frac{g_{\uparrow\uparrow}}{2}
\left|\psi_\uparrow({\bf r})\right|^4+ \frac{g_{\downarrow\downarrow}}{2}
\left|\psi_\downarrow({\bf r})\right|^4
+ g_{\uparrow\downarrow}\left|\psi_\uparrow({\bf r})\right|^2
\left|\psi_\downarrow({\bf r})\right|^2
\right] ,
\end{aligned}
\label{eq:E_tot}
\end{equation}
where $\Psi=\left(\psi_\uparrow \,\,\,
\psi_\downarrow\right)^T$ is the two-component condensate wave
function.
For simplicity, in this review we will assume $\delta=0$ and equal
intraspecies interactions $g_{\uparrow\uparrow} = g_{\downarrow\downarrow}
\equiv g$, unless otherwise specified; the effect of asymmetry of the
coupling constants will be briefly discussed at the end of the
present section. The ground-state wave function can be determined
through a variational procedure based on the ansatz \cite{Li2012PRL}
\begin{equation}
\Psi = \sqrt{\bar{n}}
\left[C_+\begin{pmatrix} \cos \theta \\ -\sin\theta \end{pmatrix}
e^{ik_1x}+C_-\begin{pmatrix} \sin \theta \\ -\cos\theta
\end{pmatrix} e^{-ik_1x} \right]
\label{eq:ansatz}
\end{equation}
where $\bar{n}=N/V$ is the average density, and $k_1$ represents
the canonical momentum where Bose--Einstein condensation takes place.
For a given value of $\bar{n}$ and $\Omega$, the variational parameters
are $C_+$, $C_-$, $k_1$ and $\theta$. Their values are determined by
minimizing the energy (\ref{eq:E_tot}) with the normalization constraint
$\int {\rm d}{\bf r} \, \Psi^\dagger\Psi=N$,
i.e., $\left|C_+\right|^2+|C_-|^2=1$.
Minimization with respect to $k_1$ yields
the general relation $2\theta=\arccos(k_1/k_0)$ fixed by
the single-particle Hamiltonian (\ref{eq:h_0_SO}).
Once the other variational parameters
are determined, it is possible to calculate key physical
quantities like, for example, the momentum distribution
accounted for by the parameter $k_1$, the total density
$n({\bf r}) = \Psi^\dagger\Psi$, the
longitudinal ($s_z({\bf r})$) and transverse ($s_x({\bf r})$,
$s_y({\bf r})$) spin densities, given by
\begin{eqnarray}
&& s_z({\bf r}) = \Psi^\dagger\sigma_z\Psi =\bar{n}
\left(|C_+|^2-|C_-|^2\right)\dfrac{k_1}{k_0} ,\label{eq:sigma_z}\\
&& s_x({\bf r}) = \Psi^\dagger\sigma_x\Psi =-\bar{n} \left[
\dfrac{\sqrt{k_0^2 -k_1^2}}{k_0} +2|C_+ C_-|\cos\left(2k_1 x+
\phi\right)\right] , \quad \label{eq:sigma_x}\\
&& s_y({\bf r}) = \Psi^\dagger\sigma_y\Psi =\bar{n}\,|C_+ C_-|
\dfrac{2 k_1}{k_0}\sin\left(2k_1 x +\phi\right) ,
\label{eq:sigma_y}
\end{eqnarray}
with $\phi$ the relative phase between $C_+$ and $C_-$,
and the corresponding spin polarizations $\langle\sigma_k\rangle =
N^{-1}\int{\rm d}{\bf r} \, s_k$ with $k=x,\,y,\,z$.
Before going on, we notice that the results (\ref{eq:sigma_x}) and
(\ref{eq:sigma_y}) hold in the spin-rotated frame where the
Hamiltonian takes the form (\ref{eq:H_N_body}).
Since the operators $\sigma_x$ and $\sigma_y$
do not commute with $\sigma_z$, in the original laboratory frame the
average value of these operators exhibits an additional oscillatory
behavior analogous to the one characterizing the laser potential
of Eq.~(\ref{eq:h_0}).

The ansatz (\ref{eq:ansatz}) exactly describes the ground state
of the single-particle Hamiltonian $h_0^{\rm SO}$ (ideal Bose gas),
reproducing all the features presented in
Sec.~\ref{subsec:Single_particle}, including the values of
the canonical momentum $k_1$. In this case the energy is independent
of $C_\pm$, reflecting the degeneracy of the ground state.

\begin{figure}[t]
\centering
\includegraphics[scale=1]{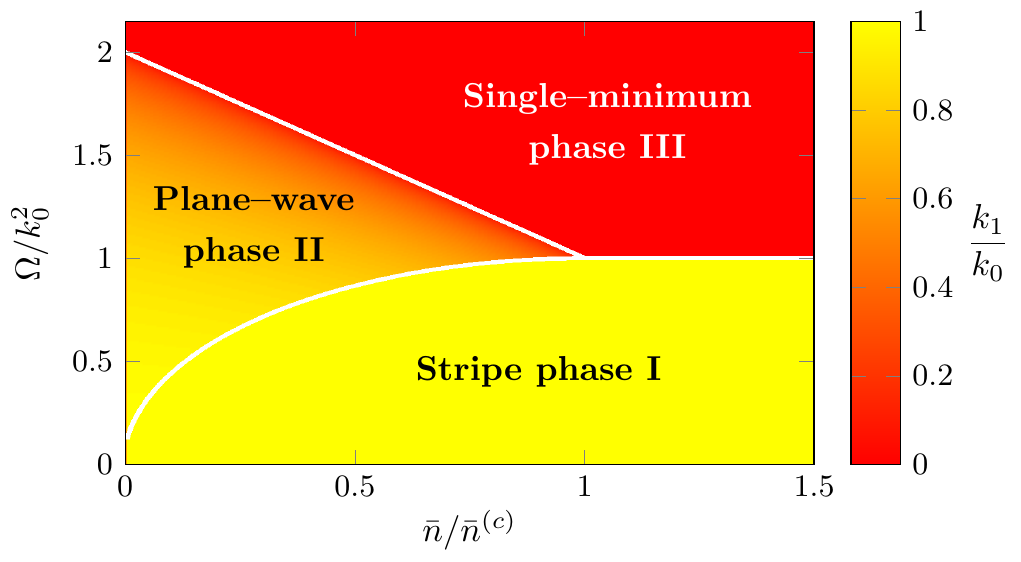}
\caption{Phase diagram of a spin-orbit-coupled BEC. The color
represents the value of $k_1/k_0$. The white solid lines identify
the phase transitions (I--II), (II--III) and (I--III). The diagram
corresponds to a configuration with $\gamma = (g -
g_{\uparrow\downarrow}) / (g + g_{\uparrow\downarrow}) = 0.0012$
consistent with the value of Ref.~\citen{Lin2011}. }
\label{fig:phase_diagram}
\end{figure}

The same ansatz is well suited also for discussing the role of
interactions, which crucially affect the explicit values of $C_+$,
$C_-$ and $k_1$. By inserting (\ref{eq:ansatz}) into (\ref{eq:E_tot}),
one finds that the energy per particle $\varepsilon=E/N$ takes
the form
\begin{equation}
\varepsilon = \frac{k_0^2}{2} -\frac{\Omega}{2 k_0} \sqrt{k_0^2-k_1^2}
-F(\beta) \frac{k_1^2}{2k_0^2} + G_1\left(1+2\beta\right) ,
\label{eq:E_per_N}
\end{equation}
where we have defined the quantities $\beta \hspace{-.6mm}
=\hspace{-.6 mm} |C_+|^2|C_-|^2 \hspace{-0.6mm} \in \hspace{-0.6mm}
[0,\,1/4]$, $G_1=\bar{n}(g+g_{\uparrow \downarrow})/4$, $G_2 =
\bar{n}(g-g_{\uparrow \downarrow})/4$ and the function
$F(\beta) = \left(k_0^2-2G_2\right)+4\left(G_1+2G_2\right)\beta$.
By minimizing (\ref{eq:E_per_N}) with respect to $\beta$ and $k_1$ we
obtain the mean-field ground state of the system.
Depending on the values of the relevant parameters $k_0$, $\Omega$,
$g$, $g_{\uparrow\downarrow}$ and $\bar{n}$, the minimum can occur
either at $k_1=0$ or at $k_1\neq0$ and $\beta$ equal to one of the
limiting values $0$ and $1/4$. Therefore, the ground state is
compatible with three distinct quantum phases; the corresponding
phase diagram is shown in Fig.~\ref{fig:phase_diagram}.

\vspace{1mm}

\textbf{(I) Stripe phase.} For small values of the Raman coupling
$\Omega$ and $g>g_{\uparrow\downarrow}$, the ground state is a
linear combination of the two plane-wave states $e^{\pm i k_1 x}$
with equal weights ($|C_+|=|C_-|=1/\sqrt{2}$), yielding a
vanishing longitudinal spin polarization
(see Eq.~(\ref{eq:sigma_z})). The most striking feature
of this phase is the appearance of density modulations in the form
of stripes according to the law
\begin{equation}
n({\bf r})=\bar{n}\left[1+\frac{\Omega}{2\left(k_0^2+G_1 \right)}
\cos\left(2k_1 x+\phi\right) \right] .
\label{eq:density_stripe}
\end{equation}
The periodicity of the fringes $\pi/k_1$ is determined by the wave vector
\begin{equation}
k_1=k_0\sqrt{1-\frac{\Omega^2}{4\left(k_0^2+ G_1\right)^2}}
\label{eq:k1_I}
\end{equation}
and differs from the one of the laser potential, equal to $\pi/k_0$
(see Eq.~(\ref{eq:h_0})).
These modulations have a deeply different nature with respect to
those exhibited by the density profile in the presence of usual
optical lattices. Indeed, they appear as the result of a spontaneous
breaking mechanism of translational invariance, with the actual
position of the fringes being given by the value of the phase $\phi$.
Because of the coexistence of BEC and crystalline order,
the stripe phase shares important analogies with supersolids
\cite{Boninsegni2012}. It also shares similarities
with the spatial structure of smectic liquid crystals.
The contrast in $n({\bf r})$ is given by
\begin{equation}
\frac{n_{\rm max}-n_{\rm min}}{n_{\rm max}+n_{\rm min}}
=\frac{\Omega}{2(k_0^2+G_1)}
\label{eq:contrast}
\end{equation}
and vanishes as $\Omega \to 0$ as a
consequence of the orthogonality of the two spin states entering
Eq.~(\ref{eq:ansatz}) (in this limit $\theta \to 0$ and $k_1
\to k_0$). It is also worth mentioning that the ansatz,
Eq.~(\ref{eq:ansatz}), for the stripe phase provides only a first
approximation which ignores higher-order harmonics caused
by the nonlinear interaction terms in the Hamiltonian.

\textbf{(II) Plane-wave phase.} For larger values of the Raman
coupling, the system enters a new phase, the so-called plane-wave
phase (also called the spin-polarized or de-mixed phase), where
Bose--Einstein condensation takes place
in a single plane-wave state with momentum ${\bf p}=k_1
\hat{\bf e}_x$ ($C_- = 0$), lying on the $x$ direction (in the
following we choose $k_1>0$). In this phase, the density is uniform
and the spin polarization is given by
\begin{equation}
\langle \sigma_z\rangle=\frac{k_1}{k_0}
\end{equation}
with
\begin{equation}
k_1=k_0\sqrt{1-\frac{\Omega^2}{4\left(k_0^2-2G_2\right)^2}} .
\label{eq:k1_II}
\end{equation}
An energetically equivalent configuration is obtained by considering
the BEC in the single-particle state with
${\bf p}=-k_1\hat{\bf e}_x$ ($C_+ = 0$). The choice between the two
configurations is determined by a mechanism of spontaneous symmetry
breaking, typical of ferromagnetic configurations.

\textbf{(III) Single-minimum phase.} At even larger values of
$\Omega$, the system enters the single-minimum phase (also called
zero-momentum phase), where the condensate has zero momentum ($k_1=
0$), the density is uniform, and the average spin polarization
$\langle \sigma_z\rangle$ identically vanishes, while $\langle
\sigma_x \rangle=-1$. Contrary to what one would naively expect,
also the single-minimum phase exhibits nontrivial properties, as we
will see in Secs.~\ref{sec:Dynamic_uniform} and
\ref{sec:Collect_trap}.

\vspace{1mm}

The chemical potential in the three phases
can be calculated from the energy per particle
(\ref{eq:E_per_N}), and takes the form
\begin{eqnarray}
\mu^{({\rm I})} & = & \;2G_1 - \frac{k_0^2 \Omega^2}{8
\left(k_0^2+ G_1 \right)^2} , \label{eq:mu_I}\\
\mu^{({\rm II})}& = & \;2 \left(G_1+G_2\right) - \frac{k_0^2
\Omega^2}{8\left(k_0^2 -2G_2\right)^2 } , \label{eq:mu_II} \\
\mu^{({\rm III})} & = & \;2G_1 +  \frac{k_0^2 - \Omega}{2} .
\label{eq:mu_III}
\end{eqnarray}
The critical values of the Rabi frequencies $\Omega$
characterizing the phase transitions can be identified
by imposing that the chemical potential
(\ref{eq:mu_I})--(\ref{eq:mu_III})
and the pressure $P=n\mu(n)-\int \mu(n)\,{\rm d}n$ be equal in the
two phases at equilibrium. The transition between the stripe and
the plane-wave phases has a first-order nature and is characterized
by different values of the densities of the two phases.
The density differences are, however, extremely small and are not
visible in Fig.~\ref{fig:phase_diagram}.
The transition between the plane-wave and the
single-minimum phases has instead a second-order nature and is
characterized by a jump in the compressibility
$n^{-1}(\partial\mu/\partial n)^{-1}$ if $G_2\neq0$ and by
a divergent behavior of the magnetic polarizability
(see Sec.~\ref{subsec:Polar_compress}).
In the low density (or weak coupling) limit, i.e.,
$G_1,\,G_2 \ll k_0^2$,
the critical value of the Raman coupling $\Omega^{({\rm I-II})}$
characterizing the transition between phases I and II is given by
the density-independent expression \cite{Ho2011,Li2012PRL}
\begin{equation}
\Omega^{({\rm I-II})} = 2 k_0^2 \sqrt{\frac{2\gamma}{1+2\gamma}} ,
\label{eq:OmegaI-II}
\end{equation}
with $\gamma=G_2/G_1$.
The transition between phases II and III instead takes place at the
higher value \cite{Li2012PRL}
\begin{equation}
\Omega^{({\rm II-III})}=2\left(k_0^2 - 2 G_2\right) ,
\label{eq:OmegaII-III}
\end{equation}
provided that the condition $\bar{n} < \bar{n}^{(c)}$ is satisfied,
with $\bar{n}^{(c)}=k_0^2/(2g\gamma)$ being the value of the density
at the tricritical point shown in Fig.~\ref{fig:phase_diagram},
where the three phases connect each other.
For higher densities one has instead a first-order
transition directly between phases I and III. We also remark
that, if $g<g_{\uparrow\downarrow}$, only phases II and III
are available, the stripe phase being always energetically
unfavorable.

The previous results can be extended to the case $\delta\neq0$ and
$g_{\uparrow\uparrow} \neq g_{\downarrow\downarrow}$. In general
one can introduce three interaction parameters:
$G_1 = \bar{n}(g_{\uparrow\uparrow} + g_{\downarrow\downarrow}
+ 2 g_{\uparrow\downarrow})/8$,
$G_2 = \bar{n}(g_{\uparrow\uparrow} + g_{\downarrow\downarrow}
- 2 g_{\uparrow\downarrow})/8$ and
$G_3 = \bar{n}(g_{\uparrow\uparrow} - g_{\downarrow\downarrow})/4$.
In the case of the states $\left|\uparrow\right\rangle =
\left|F=1,m_F=0\right\rangle$ and $\left|\downarrow\right\rangle =
\left| F=1,m_F=-1\right\rangle$ of $^{87}$Rb the values of the
scattering lengths are $a_{\uparrow\uparrow}=101.41\,a_B$ and
$a_{\downarrow\downarrow}=a_{\uparrow\downarrow}=100.94\,a_B$,
where $a_B$ is the Bohr radius. This
corresponds to $0 < G_2 = G_3/2 \ll G_1$. However, since the
differences among the scattering lengths are very small, by
properly choosing the detuning $\delta$, this effect can be
well compensated. For example, using first order perturbation
theory, one finds that the correction to the energy per
particle (\ref{eq:E_per_N}) is given, in the low density
(weak coupling) limit, by
\begin{equation}
\varepsilon^{(1)} = \left(G_3 + \frac{\delta}{2}\right)
\frac{k_1}{k_0}(\left|C_+\right|^2-\left|C_-\right|^2) .
\label{eq:E_per_N_1}
\end{equation}
By choosing $\delta = - 2 G_3$ the correction (\ref{eq:E_per_N_1})
vanishes, thus ensuring that the properties of the ground state
of the system and the transition frequencies are not affected by
the inclusion of the new terms in the Hamiltonian. If the weak
coupling condition is not satisfied, the value of $\delta$
ensuring exact compensation should depend on $\Omega$.

\begin{figure}[t]
\centering
\includegraphics[scale=1.2]{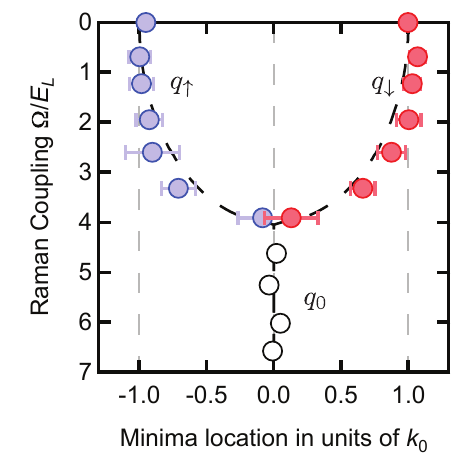}
\caption{Measured values of the canonical momentum versus $\Omega$
at $\delta=0$. The data points correspond to the minima of the
dispersion $\varepsilon_-({\bf q})$ given in
Eq.~(\ref{eq:E_single}). The Raman coupling is expressed in units of
the recoil energy $E_L = k_0^2/2$. Reprinted by permission from
Macmillan Publishers Ltd: Lin {\it et al.},
\href{http://www.nature.com}{Nature} {\bf 471}, 83-86, \copyright{}
2011.} \label{fig:Lin_k1}
\end{figure}

The emergence of a double minimum in the single-particle spectrum
and the $\Omega$ dependence of the value of $k_1$ was experimentally
observed by Lin {\it et al.} by measuring the velocity of the expanding
cloud after the release of the trap \cite{Lin2011} (see Fig.~\ref{fig:Lin_k1}).
The double-minimum structure vanishes at the
predicted value (\ref{eq:OmegaII-III}) of the Raman coupling giving
the transition between the plane-wave and the single-minimum phases.
In the same experiment, at a lower value of $\Omega$, they identified
another transition between a mixed phase, characterized by two
different canonical momentum components overlapping in space,
and a de-mixed phase, where the two components coexist but are
spatially separated. The critical Raman coupling at which the latter
transition has been observed is in good agreement with the prediction
$\Omega^{({\rm I-II})} = 0.19\,E_r$ for the transition
frequency between the stripe and the plane-wave phases, obtained from
Eq.~(\ref{eq:OmegaI-II}) with the $^{87}$Rb value $\gamma=0.0012$.
However, it has not been possible to observe directly the density
modulations because of the smallness of their contrast and
periodicity (see Sec.~\ref{sec:Excitation_stripe}).

Finally, we mention that the critical density $\bar{n}^{(c)}$ is
very large in the experimental conditions of Ref.~\citen{Lin2011}, thus
preventing the access to the regime where the first-order transition
between the stripe and the single-minimum phases takes place. A
strong reduction of the value of $\bar{n}^{(c)}$ could be
achieved, for example, by considering configurations where the
interspecies coupling strength $g_{\uparrow\downarrow}$ is
significantly smaller than the intraspecies ones
$g_{\uparrow \uparrow}$, $g_{\downarrow \downarrow}$,
as discussed in Sec.~\ref{subsec:Exp_stripes}.

\subsection{Magnetic polarizability and compressibility}
\label{subsec:Polar_compress}
As already pointed out, the transition between the plane-wave and
the single-minimum phases is characterized by a divergent behavior
of the magnetic polarizability $\chi^{}_M$. This quantity is defined as
the linear response $\chi^{}_M = (\langle h|\sigma_z|h\rangle -
\langle h=0|\sigma_z|h=0\rangle)/h$
to a static perturbation of the form
$- h\sigma_z$, and can be calculated by generalizing the ground-state
condensate wave function (\ref{eq:ansatz}) to include the presence of
a small magnetic field $h$. In the plane-wave and the
single-minimum phases, the magnetic polarizability takes the simple form
\cite{Li2012EPL}
\begin{eqnarray}
\chi^{({\rm II})}_M &=& \frac{\Omega^2}{\left(k_0^2 - 2G_2\right)
\left[4 \left(k_0^2- 2G_2\right)^2-\Omega^2\right]} ,
\label{eq:chi_M_II}\\
\chi^{({\rm III})}_M &=& \frac{2}{\Omega-2\left(k_0^2 - 2 G_2
\right)} ,
\label{eq:chi_M_III}
\end{eqnarray}
and exhibits a divergent behavior at the transition
between the two phases. Indeed, when approaching the transition
(\ref{eq:OmegaII-III}) from above or below, the values of
$\chi^{}_M$ differ by a factor 2, revealing the second-order nature
of the phase transition \cite[\S 144]{bookLandau1980}. It is worth pointing
out that, if $G_2=0$, the
calculation of $\chi^{}_M$ reduces to the ideal gas value, which is
found to be related to the effective mass (\ref{eq:mass_12}) and
(\ref{eq:mass_3}) by the simple relation
\begin{equation}
\frac{m^\ast}{m\;} = 1+k_0^2\,\chi^{}_M .
\label{eq:chi_mass}
\end{equation}
The divergent behavior of the magnetic polarizability near the
second-order phase transition was experimentally confirmed by Zhang
{\it et al.} through the study of the center-of-mass oscillation
\cite{Zhang2012} (see also the discussion in Sec.~\ref{sec:Collect_trap}).
Concerning the stripe phase, the calculation
of $\chi^{}_M$ yields a complicated expression which, in the weak
coupling limit $G_1,\,G_2 \ll k_0^2$, reduces to the simplified form
\begin{equation}
\chi^{({\rm I})}_M = \frac{\Omega^2 - 4 k_0^4}
{\left(G_1+2G_2\right)\Omega^2 - 8 G_2 k_0^4} .
\label{eq:chi_M_I}
\end{equation}
Notice that $\chi^{({\rm I})}_M$ diverges at the critical frequency
providing the transition to the plane-wave phase (see Eq.~(\ref{eq:OmegaI-II})).
However, Eq.~(\ref{eq:chi_M_I}) is valid only in the weak coupling limit,
and the inclusion of higher-order terms makes the value of
$\chi^{}_M$ finite at the transition.

The thermodynamic compressibility
$\kappa_T=n^{-1}(\partial\mu/\partial n)^{-1}$ in all the phases
can be calculated from the expressions of the chemical potential
(see Eqs.~(\ref{eq:mu_I})--(\ref{eq:mu_III})),
\begin{eqnarray}
&& 1/\kappa^{({\rm I})}_T  = 2G_1+\frac{G_1 k_0^2
\Omega^2}{4\left(k_0^2+G_1\right)^3} ,\label{eq:kappa_T_I}\\
&& 1/\kappa^{({\rm II})}_T = 2\left(G_1+G_2\right)-\frac{G_2 k_0^2
\Omega^2}{2\left(k_0^2-2G_2\right)^3} ,\label{eq:kappa_T_II}
\vspace{1.mm}\\
&& 1/\kappa^{({\rm III})}_T = 2 G_1 .\label{eq:kappa_T_III}
\end{eqnarray}
For an interacting Bose gas, the compressibility
(\ref{eq:kappa_T_I})--(\ref{eq:kappa_T_III}) has always a finite value. It
is discontinuous at the first-order transition between the stripe and the
plane-wave phases; furthermore, if $G_2 \neq 0$, it exhibits a jump also
at the second-order transition between the plane-wave and the
single-minimum phases. However, as we will show in the next section,
the sound velocity is continuous across the latter transition.

\section{Dynamic Properties of the Uniform Phases}
\label{sec:Dynamic_uniform}
Spin-orbit coupling affects in a deep way also the dynamic behavior of
a BEC, giving rise to exotic features in the excitation spectrum,
such as the emergence of a rotonic structure when
one approaches the transition from the plane-wave to the
stripe phase \cite{Zheng2012,Martone2012}, the suppression of the sound
velocity near the transition between the plane-wave and the
single-minimum phases \cite{Martone2012}, a double gapless band
structure in the stripe phase \cite{Li2013}, etc.
In the present section and in the next one,
we focus on the dynamic behavior of a spin-orbit-coupled BEC in the
uniform phases II and III. The properties of the stripe phase will
be discussed in Sec.~\ref{sec:Excitation_stripe}.

\subsection{Dynamic density response function. Excitation spectrum}
\label{subsec:Bogoliubov_uniform}

In order to investigate the dynamic properties of a
spin-orbit-coupled BEC it is useful to evaluate its dynamic density
response function. This can be done by adding a time-dependent
perturbation $V_{\lambda} = -\lambda e^{i({\bf q}\cdot{\bf r}-
\omega t)}+\rm{H.c.}$ to the single-particle Hamiltonian
(\ref{eq:h_0_SO}). The direction of the wave vector ${\bf q}$ is
characterized by the polar angle  $\alpha \in [0,\,\pi]$ with
respect to the $x$ axis.
The density response function is then calculated through
the usual definition
$\chi({\bf q},\omega) = \lim_{\lambda \to 0}\delta n_{\bf q} /
(\lambda e^{-i\omega t})$, where $\delta n_{\bf q}$ are the
fluctuations of the ${\bf q}$ component of the density induced by
the external perturbation.\footnote{The spin-density response function
can be calculated with an analogous procedure by adding a perturbation
$\sigma_z V_\lambda$ to (\ref{eq:h_0_SO}).} In the following we derive
$\chi({\bf q},\omega)$ by solving the time-dependent
Gross--Pitaevskii equation
\begin{equation}
i \frac{\partial \Psi}{\partial t} = \left[h_0^{\rm SO} +
V_\lambda + \frac{1}{2}\left(g+g_{\uparrow\downarrow}\right)
\left(\Psi^\dagger \Psi\right)
+ \frac{1}{2}\left(g-g_{\uparrow\downarrow}\right)
\left(\Psi^\dagger \sigma_z \Psi\right)\sigma_z\right]\Psi ,
\label{eq:spinor_GP}
\end{equation}
where $h_0^{\rm SO}$ is the single-particle Hamiltonian
(\ref{eq:h_0_SO}) with $\delta=0$. Since we are focusing on
phases II and III, where the ground-state density is uniform,
the spinor wave function $\Psi$ in Eq.~(\ref{eq:spinor_GP})
can be written in the simple form
\begin{equation}
\Psi=  e^{- i \mu t} \left[\sqrt{\bar{n}} \begin{pmatrix} \cos\theta\\
-\sin\theta \end{pmatrix} e^{i k_1 x} + \begin{pmatrix}
u_{\uparrow}({\bf r}) \\ u_{\downarrow}({\bf r})
\end{pmatrix} e^{-i \omega t} + \begin{pmatrix} v^\ast_{\uparrow}
({\bf r}) \\ v^\ast_{\downarrow}({\bf r}) \end{pmatrix} e^{i
\omega t}\right] .
\label{eq:spinor_uniform}
\end{equation}
The terms depending on the Bogoliubov amplitudes $u$ and $v$
provide the deviations in the order parameter with respect to
equilibrium, caused by the external perturbation. In the linear
(small $\lambda$) limit we find the result (near the poles one should
replace $\omega$ with $\omega +i0$)
\begin{equation}
\chi({\bf q},\omega)= \frac{- N q^2\left[ \omega^2 - 4 k_1 q
\cos\alpha \, \omega + a(q,\alpha)\right]} {\omega^4 -4 k_1 q
\cos\alpha \, \omega^3 + b_2(q,\alpha)\, \omega^2 + k_1 q
\cos\alpha \, b_1(q,\alpha)\, \omega + b_0(q,\alpha)} ,
\label{eq:response_uniform}
\end{equation}
where the coefficients $a$ and $b_i$ are even
functions of $q \equiv|{\bf q}|$ and $\cos\alpha$, implying that
$b_i(q,\alpha) = b_i(q,\pi\pm\alpha)$ (the same for $a$), and
their actual values depend on whether one is in phase II or III (see
App.~\ref{app:Response_coeff}). In the plane-wave phase, the odd terms
in $\omega$ entering the response function reflect the
lack of parity and time-reversal symmetry of the ground-state wave
function; in the single-minimum phase, however, one has $k_1=0$
and thus the symmetry is restored.

Notice that the response function
(\ref{eq:response_uniform}) reduces to the usual Bogoliubov form
$\chi({\bf q},\omega)= - N q^2/[\omega^2 - q^2(2G_1 +q^2/4)]$
when $G_2 =0$ and $\Omega =0$, characterizing the response of a BEC
gas in the absence of spin-orbit coupling. It is also worth pointing
out that, since $V_\lambda$ commutes with the unitary transformation
yielding the Hamiltonian in the spin-rotated frame
(see Sec.~\ref{subsec:Single_particle}), the expression for
$\chi({\bf q},\omega)$ is the same as in the original laboratory
frame, and thus all the results based on the calculation in the
spin-rotated frame are relevant for actual experiments.

\begin{figure}
\centering
\includegraphics[scale=1]{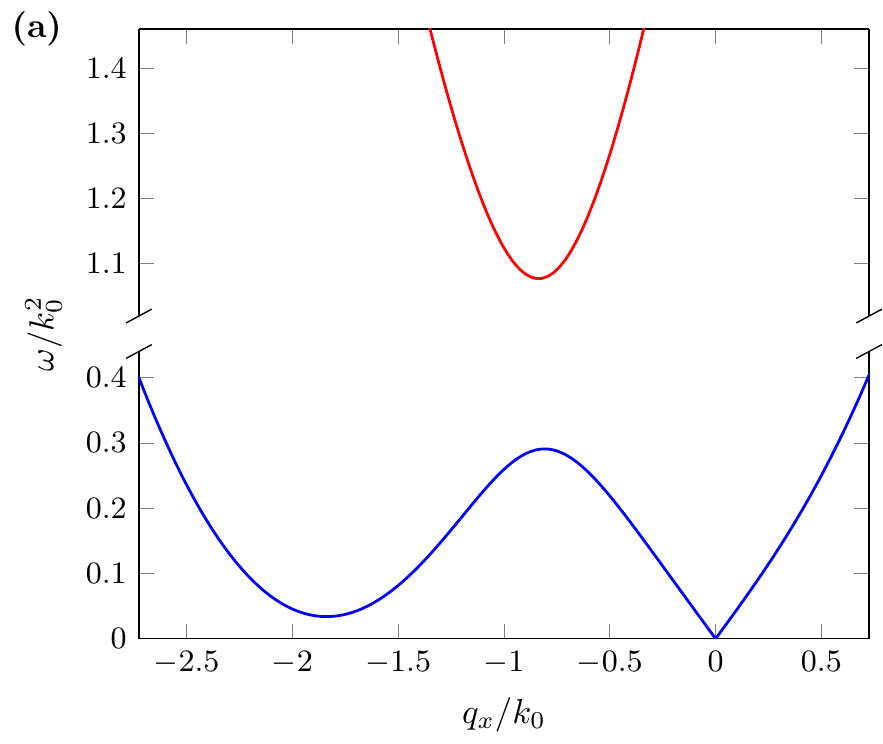}\\
\includegraphics[scale=1]{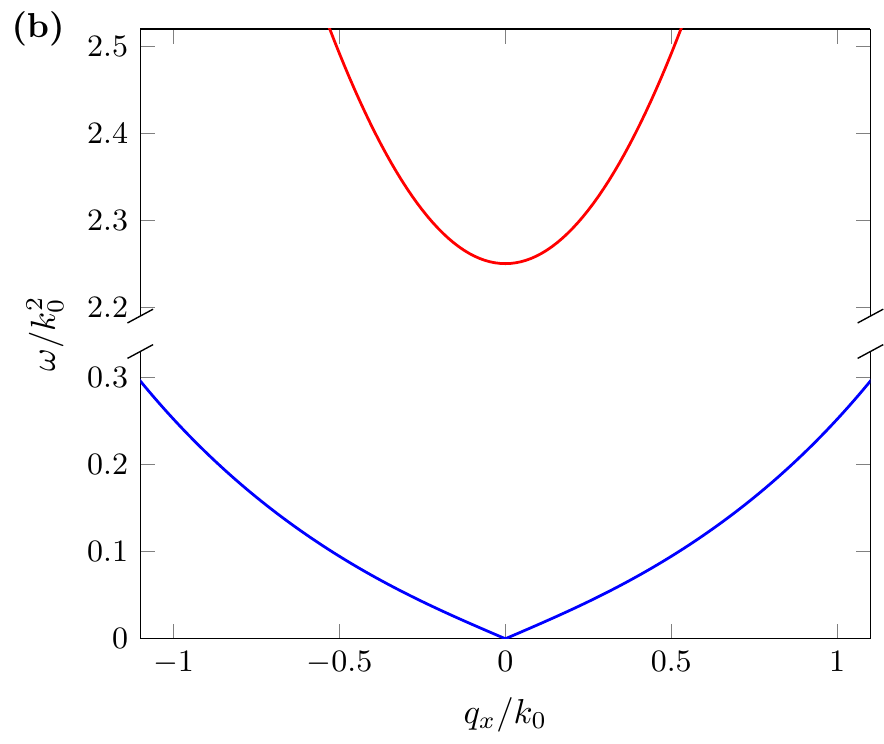}
\caption{Excitation spectrum {\bf (a)} in phase II
($\Omega/k_0^2=0.85$) and {\bf (b)} in phase III
($\Omega/k_0^2=2.25$) as a function of $q_x$ ($q_y=q_z=0$),
calculated in the experimental conditions of Ref.~\citen{Zhang2012}.
The blue and red lines represent the lower and upper branches,
respectively. In phase II the spectrum is not symmetric and exhibits
a roton minimum for negative $q_x$, whose energy becomes smaller and
smaller as one approaches the transition to the stripe phase at
$\Omega/k_0^2=0.095$. The other parameters: $G_1/k_0^2=0.12$,
$\gamma = G_2/G_1= 10^{-3}$.} \label{fig:disp_uniform}
\end{figure}

The frequencies of the elementary excitations are given by the poles
of the response function $\chi$, i.e., by the zeros of
\begin{equation}
\omega^4 - 4 k_1 q \cos\alpha \, \omega^3 + b_2 \,\omega^2 + k_1 q
\cos\alpha \, b_1 \,\omega + b_0 = 0 .
\label{eq:disp_uniform}
\end{equation}
The solutions of this equation provide two separated branches, as
shown in Fig.~\ref{fig:disp_uniform}(a) and (b) for phase II and phase
III respectively. The lower one is gapless and exhibits a phonon
dispersion at small $q$, while the upper one is gapped as a
consequence of the Raman coupling. For example, in phase III the
gap between the two branches is given, at ${\bf q}=0$,
by $\Delta = \sqrt{\Omega(\Omega+4G_2)}$. Differently from the
single-minimum phase, the excitation spectrum in the plane-wave
phase is not symmetric under inversion of $q_x$ into $-q_x$, as
a consequence of the symmetry-breaking terms appearing in
Eq.~(\ref{eq:response_uniform}).
For negative values of $q_x$, the lower branch in phase II exhibits
a very peculiar feature, resulting in the
emergence of a roton minimum, which becomes more and more pronounced
as one approaches the transition to the stripe phase. The occurrence
of the rotonic structure in spin-orbit-coupled BECs shares
interesting analogies with the case of dipolar gases in quasi-2D
configurations \cite{Santos2003} and of condensates with soft-core,
finite-range interactions \cite{Macri2013,Saccani2012}. The physical
origin of the roton minimum is quite clear. In phase II the ground
state is twofold degenerate, and it is very favorable for atoms to
be transferred from the BEC state with momentum ${\bf p} =
k_1 \hat{\bf e}_x$ to the empty state at ${\bf p} = -k_1
\hat{\bf e}_x$. The excitation spectrum has been recently measured
using Bragg spectroscopy techniques, confirming the occurrence
of a characteristic rotonic structure \cite{Ji2014,Khamehchi2014}
(see also Ref.~\citen{Ha2014} for the case of shaken optical
lattices).\footnote{In the experiments of
Refs.~\citen{Ha2014,Ji2014,Khamehchi2014} the excitation spectrum
has been measured on top of the BEC state with
momentum ${\bf p} = - k_1 \hat{\bf e}_x$, for which the roton minimum,
differently from the case discussed above, appears at positive values of
$q_x$.}

\subsection{Static response function and static structure factor}
\label{subsec:Structure_uniform}

The static response function $\chi({\bf q}) \equiv
\chi({\bf q},\omega=0)/N$ can be derived directly from
Eq.~(\ref{eq:response_uniform}). Its $q=0$ value
$\mathcal{K}\equiv \chi(q=0)$ is given by
\begin{eqnarray}
\mathcal{K}^{-1}_{\rm II} & = & 2 G_1 + \frac{2 G_2
k_1^2\left(k_1^2\cos^2\alpha + k_0^2\sin^2\alpha - 2
G_2\right)}{k_1^2\left(k_0^2\cos^2\alpha - 2 G_2\right) +
k_0^4\sin^2\alpha} ,
\label{eq:compress_II}\\
\mathcal{K}_{\rm III}^{-1} & = & 2 G_1
\label{eq:compress_III}
\end{eqnarray}
in the plane-wave phase II and the single-minimum phase III,
respectively. The anisotropy of $\mathcal{K}$ in phase II caused
by the spin interaction term $G_2$ is revealed by the last term
of Eq.~(\ref{eq:compress_II}) which depends on the polar angle $\alpha$.
It is also worth pointing
out that $\mathcal{K}_{\rm II}$ coincides with the thermodynamic
compressibility $\kappa^{({\rm II})}_T$ (see Eq.~(\ref{eq:kappa_T_II}))
only along the $x$ direction, i.e., when $\sin\alpha=0$. In this case,
$\mathcal{K}$ also exhibits a jump at the transition between phases
II and III. This marks a difference with respect to the behavior of
the frequencies of the elementary excitations, fixed by
Eq.~(\ref{eq:disp_uniform}), which are always continuous functions of
$\Omega$ at the transition for all values of ${\bf q}$.

\begin{figure}
\centering
\includegraphics[scale=1]{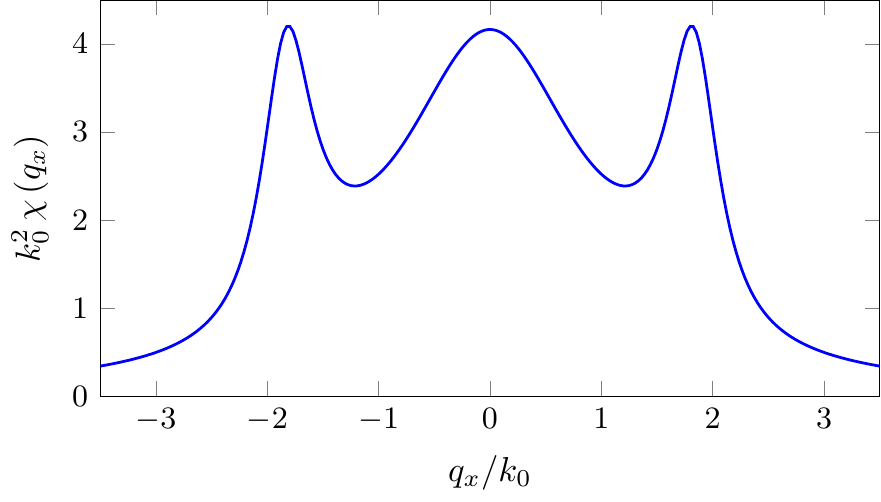}
\caption{Static response in phase II as a function of $q_x$
($q_y=q_z=0$). The curve is symmetric and exhibits a typical peak
near the roton momentum. The parameters are $\Omega/k_0^2=0.85$,
$G_1/k_0^2=0.12$ and $\gamma = G_2/G_1= 10^{-3}$.}
\label{fig:pw_statrespfunc}
\end{figure}

Far from the phonon regime, the occurrence of the roton minimum is
reflected in an enhancement in the static response function
$\chi(q_x)$ close to the roton momentum, as shown in
Fig.~\ref{fig:pw_statrespfunc}, representing a typical tendency
of the system towards crystallization. When the roton frequency
vanishes, $\chi(q_x)$ exhibits a divergent behavior.
A simple analytic expression for the corresponding value of
the Raman coupling $\Omega$ is obtained in the weak coupling limit
$G_1, G_2 \ll k^2_0$, where we find that the critical value exactly
coincides with the value (see Eq.~(\ref{eq:OmegaI-II})) characterizing the
transition between the plane-wave and the stripe phases. For larger
values of the coupling constants $G_1$ and $G_2$, the critical value
takes place for values of the Raman coupling smaller than the value
at the transition, exhibiting the typical spinoidal behavior of
first-order liquid-crystal phase transitions.

\begin{figure}
\centering
\includegraphics[scale=1]{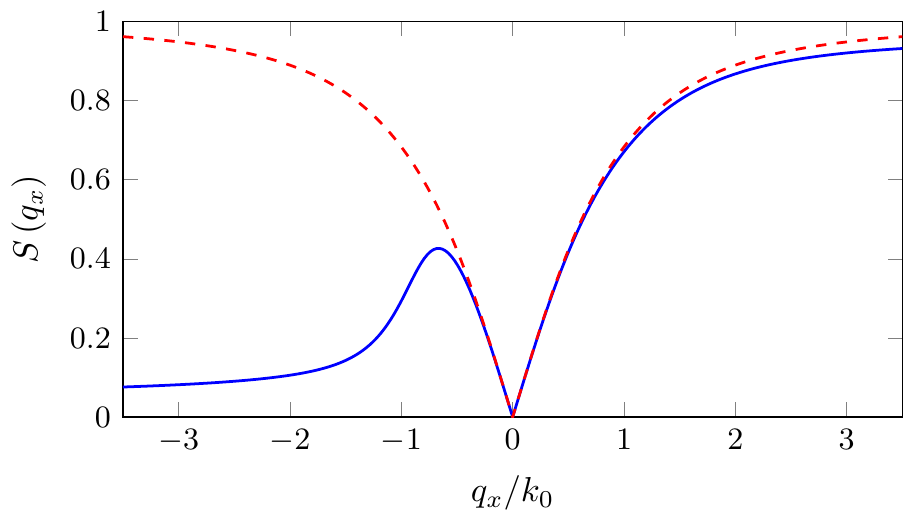}
\caption{Contribution of the lower branch to the static structure
factor in phase II as a function of $q_x$ (blue solid line),
compared with the total $S(q_x)$ (red dashed line). The parameters
are $\Omega/k_0^2=0.85$, $G_1/k_0^2=0.12$ and $\gamma = G_2/G_1=
10^{-3}$.} \label{fig:pw_statstrucfact}
\end{figure}

The dynamic structure factor at $T=0$ can be calculated from the
response function (\ref{eq:response_uniform}) through the relation
$S({\bf q},\omega) = \pi^{-1}{\rm Im}\, \chi({\bf q},
\omega)$ for $\omega \ge 0$ and $S({\bf q},\omega)= 0$ for
negative $\omega$. In the plane-wave phase, the condition ${\rm Im}
\, \chi({\bf q},\omega)= -{\rm Im}\, \chi(-{\bf q},
-\omega)$, characterizing the imaginary part of the response
function, is still satisfied, but the symmetry relation ${\rm Im}\,
\chi({\bf q},\omega) = {\rm Im}\, \chi(-{\bf q},\omega)$ is
not ensured, and consequently one finds $S({\bf q},\omega)\ne
S(- {\bf q},\omega)$. This affects several well-known
equalities involving sum rules, which have to be formulated
in a more general way to account for the breaking of inversion
symmetry in the plane-wave phase. An example is the $f$-sum rule
$\int {\rm d}\omega\, [S({\bf q},\omega)+ S(-{\bf q},
\omega)]\,\omega = Nq^2$, which is exactly satisfied,
as one can deduce from the correct large $\omega$ behavior of
the density response function:
$\chi({\bf q},\omega)_{\omega \to \infty} = - N q^2/\omega^2$
\cite{bookPitaevskii2003}. On the other hand, the inversion
symmetry of the static structure factor
$S({\bf q})=\int_0^\infty {\rm d}\omega \,S({\bf q},\omega) /N$
is always ensured, since it is a general feature following
from the completeness relation and the commutation relation
involving the density operators: $S({\bf q})-S(-{\bf q}) =
\langle\, [\rho_{\bf q},\rho_{-{\bf q}}]
\,\rangle = 0$.

It is worth pointing out that, despite the strong enhancement
exhibited by the
static response function $\chi(q_x)$, the static structure factor
$S(q_x)$ does not exhibit any peaked structure near the roton point.
This is different from what happens, for example, in superfluid
helium.\footnote{At finite temperature $T$ one instead expects the
static structure factor to be significantly peaked near the roton
minimum, provided the roton energy  is small compared to $T$, as a
consequence of the thermal excitations of rotons, similarly to what
is predicted for quasi-2D dipolar gases \cite{Klawunn2011}.} In
Fig.~\ref{fig:pw_statstrucfact} we show the static structure factor
$S(q_x)$ together with the contribution to the integral $S(q_x) =
\int {\rm d}\omega \,S(q_x,\omega)/N$ arising from the lower
branch of the elementary excitations. The figure shows that the
lower-branch contribution is not symmetric for exchange of $q_x$
into $-q_x$, even if the total $S(q_x)$ is symmetric, as we have
showed previously. Remarkably, the
strength carried by the lower branch is significantly peaked for
intermediate values of $q_x$ between the phonon and the roton
regimes, in the so-called maxon region, where the lower branch
of the excitation spectrum exhibits a maximum.

\subsection{Velocity and density vs spin nature of the sound mode}
\label{subsec:sound_velo_uniform}

The low-frequency excitations at small $q$, i.e., the sound waves,
can be easily obtained by setting $\omega=c q$, where
$c$ is the sound velocity, and keeping the leading terms
proportional to $q^2$ in Eq.~(\ref{eq:disp_uniform}). This allows us to
obtain the sound velocity in the plane-wave and the single-minimum
phases,
\begin{align}
c_{\rm II} = &{}\, \dfrac{1}{k_0^4 - 2 G_2 k_1^2}
\bigg\{\begin{aligned}[t] & G_2 k_1\left(k_0^2
- k_1^2 \right)\cos \alpha \\[2mm]
&\hspace*{-3cm} +\sqrt{ 2\left[G_1 k_0^4 + G_2 k_1^2 \left(k_0^2 - 2 G_1 -
2 G_2\right)\right] \left[k_0^4 - 2 G_2 k_1^2 - k_0^2 \left(k_0^2 -
k_1^2\right) \cos^2\alpha \right]}\bigg\} ,
\end{aligned}
\label{eq:sound_velo_II} \\
c_{\rm III} = &{}\, \sqrt{2 G_1\left(1 - \dfrac{2 k_0^2 \cos^2
\alpha}{\Omega + 4 G_2}\right)} .\label{eq:sound_velo_III}
\end{align}
Approaching the transition between the two phases, both sound
velocities
exhibit a strong reduction along the $x$ direction ($\cos\alpha=\pm
1$), caused by the spin-orbit coupling. This suppression can be
understood in terms of the increase of the effective mass associated
with the single-particle dispersion (\ref{eq:E_single}). At the
transition, where the velocity of sound modes propagating along the
$x$ direction vanishes, the elementary excitations exhibit a
different $q^2$ dependence. On the other hand, the sound velocities
along the other directions ($\alpha \neq 0$, $\alpha \neq \pi$) remain
finite
at the transition. The sound velocity in phase II shows a further
interesting feature caused by the lack of parity symmetry. The
asymmetry effect in $c_{\rm II}$ is due to the presence of the
first term in the numerator of Eq.~(\ref{eq:sound_velo_II}), therefore
the symmetry will be recovered if $G_2=0$ or $\alpha=\pi/2$
(corresponding to phonons propagating along the directions orthogonal
to the $x$ axis).

\begin{figure}[t]
\centering
\includegraphics[scale=1]{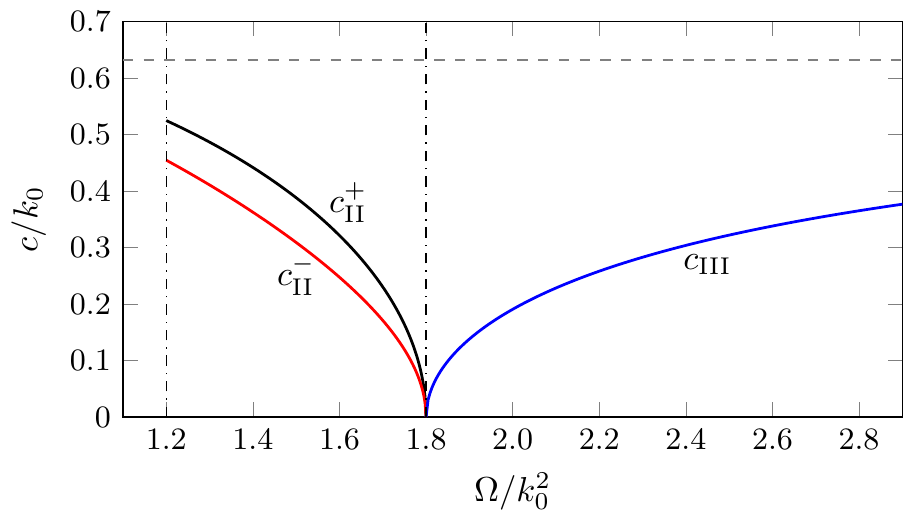}
\caption{Sound velocity as a function of the Raman coupling for the
following choice of parameters: $G_1/k_0^2 = 0.2$, $G_2/k_0^2 =
0.05$. The two sound velocities in phase II correspond to phonons
propagating in the direction parallel ($c^+_{\rm II}$) and
antiparallel ($c^-_{\rm II}$) to $k_1$. The horizontal dashed line
corresponds to the value $\sqrt{2G_1}=0.63\,k_0$ of the sound
velocity in the absence of spin-orbit and Raman coupling. The
vertical dash-dotted lines indicate the critical values of $\Omega$
at which the I-II and II-III phase transitions take place.}
\label{fig:sound_velo_uniform}
\end{figure}

The role played by the spin degree of freedom in the propagation of
the sound can be better understood by relating the sound velocity to
the magnetic polarizability $\chi^{}_M$
(see Eqs.~(\ref{eq:chi_M_II}) and (\ref{eq:chi_M_III}))
and the $q=0$ static response $\mathcal{K}$
(see Eqs.~(\ref{eq:compress_II}) and (\ref{eq:compress_III})).
One finds the result
\begin{equation}
c(\alpha) c(\alpha+\pi) = \frac{1+k_0^2 \,\chi^{}_M \sin^2
\alpha}{\mathcal{K}\left(1+k_0^2\,\chi^{}_M \right)} ,
\label{eq:c+_c-}
\end{equation}
holding in both phases II and III. The above equation generalizes the
usual relation $c^2= 1/\mathcal{K} = n(\partial\mu/ \partial n)$
between the sound velocity and the compressibility holding in usual
superfluids. It explicitly shows that, along the $x$ direction,
where $\sin \alpha =0$, the sound velocity $c$ vanishes at the
transition because of the divergent behavior of the magnetic
polarizability. The sound velocity along the $x$ axis as a function
of $\Omega$ is shown in Fig.~\ref{fig:sound_velo_uniform} for a
configuration with relatively large $G_2$, emphasizing the
difference between $c^+_{\rm II} = c_{\rm II}(\alpha=0)$
and $c^-_{\rm II} = c_{\rm II}(\alpha = \pi)$. The
suppression effect exhibited by the sound velocity near the II-III
phase transition is particularly remarkable in the single-minimum
phase III where BEC takes place in the ${\bf p}=0$ state and the
compressibility of the gas is unaffected by spin-orbit coupling. It
explicitly reveals the mixed density and spin nature of the sound
waves, with the spin nature becoming more and more important as one
approaches the phase transition where $\chi^{}_M$ diverges.

The combined density and spin nature of sound waves is also nicely
revealed by the relative amplitudes of
the density $\delta n$ and spin-density $\delta s$
oscillations in the $q \to 0$ limit, characterizing the
propagation of sound. In terms of the magnetic polarizability
$\chi^{}_M$ we find
\begin{alignat}{3}
\left(\frac{\delta s}{\delta n}\right)_{\rm II}&={}&&{}
\frac{\sqrt{1 + \left(k_0^2 -2 G_2 \right) \chi^{}_M}}{1+k_0^2\,
\chi^{}_M} + \frac{k_0\, \chi^{}_M \cos \alpha}{1+ k_0^2\,\chi^{}_M}
\sqrt{ \frac{2 \left[ G_2 +G_1 \left(1+k_0^2\,\chi^{}_M \right)
\right]}{1 +k_0^2\,\chi^{}_M \sin^2 \alpha}} ,
\label{eq:amplitude_II}\\
\left(\frac{\delta s}{\delta n}\right)_{\rm III}&={}&&
\frac{2 k_0\,\chi^{}_M \cos\alpha \, \sqrt{G_1}}{\sqrt{2\left(1+k_0^2\,
\chi^{}_M \right) \left(1+k_0^2\,\chi^{}_M \sin^2\alpha \right)}}
\label{eq:amplitude_III}
\end{alignat}
in the plane-wave and the single-minimum phases respectively. The above
equations show that, near the transition between phases II and III,
the amplitude of the spin-density fluctuations $\delta s$ of the
sound waves propagating along the $x$ direction ($\sin\alpha=0$) is
strongly enhanced with respect to the density fluctuations $\delta
n$, as a consequence of the divergent behavior of the magnetic
polarizability. In particular, very close to the phase transition the
relative amplitude is given by
\begin{equation}
\frac{\delta s}{\delta n} \sim \sqrt{2 G_1 \chi^{}_M}
\label{eq:amplitude_trans}
\end{equation}
in both phases II and III.
This suggests that an effective way to excite these
phonon modes near the transition is through a coupling with the spin
degree of freedom as recently achieved in two-photon Bragg
experiments on Fermi gases \cite{Hoinka2012}. For sound waves
propagating in the direction orthogonal to $x$ the situation is
instead different. In particular in phase III sound waves are purely
density oscillations ($\delta s=0$).

It is finally interesting to understand the role played by
the sound waves in terms of sum rules. The phonon mode exhausts
the compressibility sum rule
$\int {\rm d} \omega \,[S({\bf q},\omega)+S(-{\bf q},\omega)]
/\omega$ at small $q$, as one can easily prove from
Eq.~(\ref{eq:response_uniform}).
However, different from ordinary superfluid, it gives only a small
contribution to the $f$-sum rule as one approaches the second-order
transition. This contribution becomes vanishingly small at the
transition for wave vectors ${\bf q}$ oriented along the $x$
direction. Also, the static structure factor $S({\bf q})$ is
strongly quenched compared to usual BECs. This results in an
enhancement of the quantum fluctuations of the order parameter,
as predicted by the uncertainty principle inequality
\cite{Pitaevskii1991,Pitaevskii1993}. This effect is, however,
small because the sound velocity vanishes only along the $x$
direction \cite{Li2012PRL}.

\section{Collective Excitations in the Trap}
\label{sec:Collect_trap}

\begin{figure}
\centering
\includegraphics[scale=1]{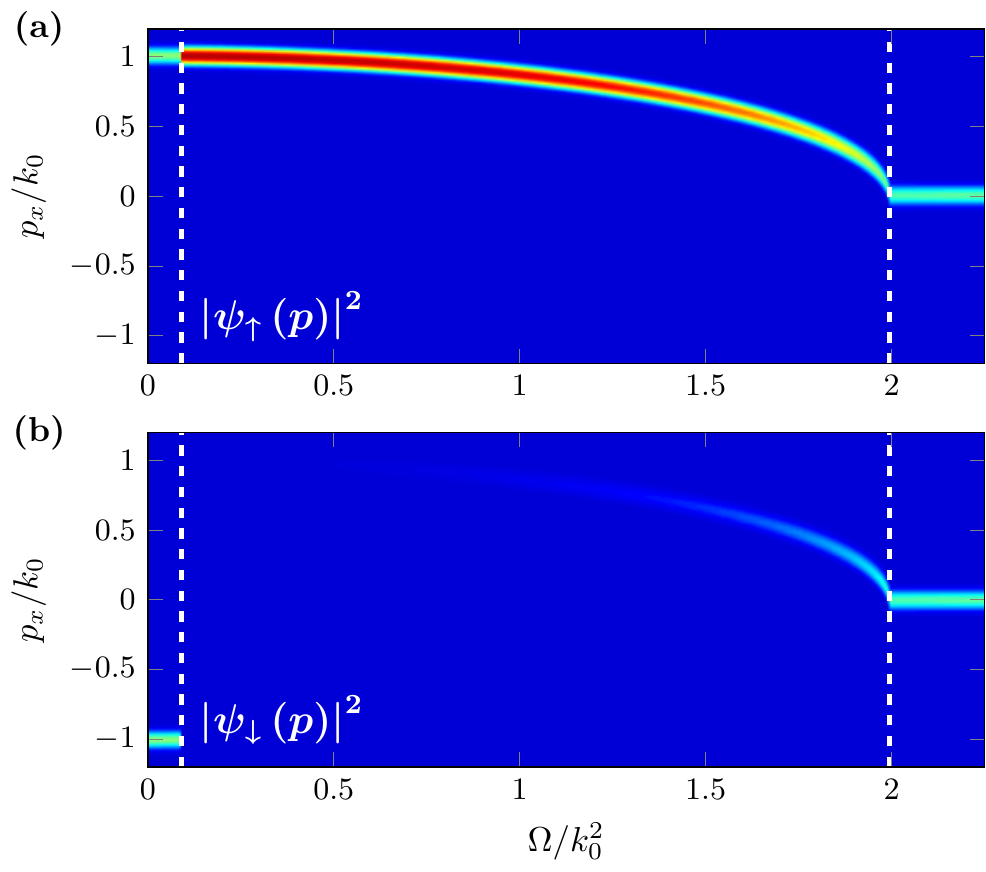}
\includegraphics[scale=1]{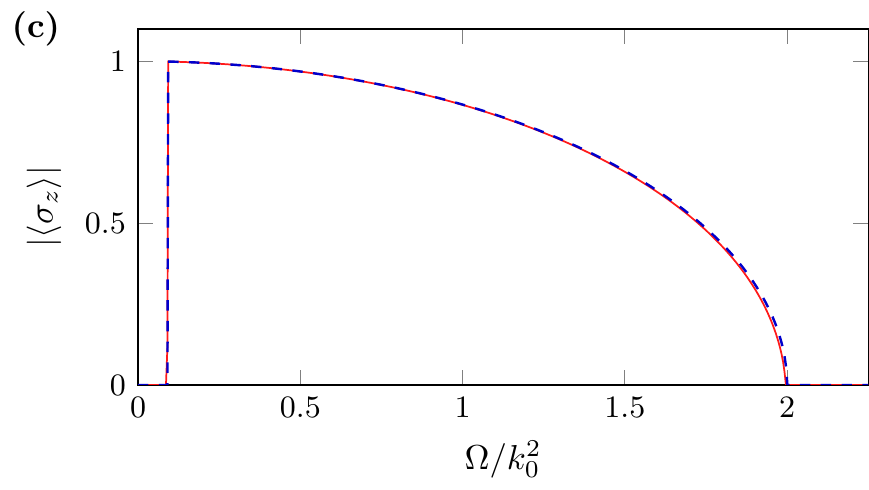}
\caption{{\bf (a-b)}. Momentum distribution for the two spin
components as a function of $\Omega$. The white dashed lines
indicate the transition frequencies calculated using
(\ref{eq:OmegaI-II}) and (\ref{eq:OmegaII-III}). {\bf (c)}. Spin
polarization $|\langle\sigma_z \rangle| = |N_{\uparrow}-
N_{\downarrow}| / N$ as a function of $\Omega$ in the trapped case
(red solid line) and in the uniform case using the density in the
center of the trap (blue dashed line). The parameters are chosen as
follows: $\omega_x=2\pi\times 40\,$Hz, $\omega_x/k_0^2 = 0.01$,
$\delta=0$, $g_{\uparrow\uparrow}= g_{\downarrow \downarrow} = 4\pi
\times 101.20\, a_B$, $g_{\uparrow\downarrow}= 4\pi\times 100.94
\,a_B$, where $a_B$ is the Bohr radius. The density in the center of
the trap corresponds to $n\simeq 1.9\times 10^{13}\,$cm$^{-3}$. }
\label{fig:psi_p_sigmaz}
\end{figure}

In this section we discuss the collective excitations for a
harmonically trapped BEC with spin-orbit coupling. First one should
notice that, in typical experimental conditions, the spin-orbit
coupling strength, usually quantified by the recoil energy
$E_r=k_0^2/2$, is much larger than the trapping frequencies.
As a consequence, one expects that the three phases occurring
in uniform matter due to the spin-orbit coupling survive
also in the presence of harmonic trapping.
This can be verified by solving numerically the
Gross--Pitaevskii equation
\begin{equation}
i \frac{\partial \Psi}{\partial t} = \left[h_0^{\rm SO} +
V_{\rm ext}({\bf r}) + \frac{1}{2}\left(g+g_{\uparrow\downarrow}
\right) \left(\Psi^\dagger \Psi\right) + \frac{1}{2} \left(g-
g_{\uparrow\downarrow} \right) \left(\Psi^\dagger \sigma_z
\Psi\right)\sigma_z\right]\Psi
\label{eq:GPE}
\end{equation}
for the condensate wave function, with $V_{\rm ext}({\bf r})=
(\omega_x^2x^2 + \omega_y^2 y^2 + \omega_z^2 z^2) /2$ representing
the external trapping potential. Figure~\ref{fig:psi_p_sigmaz} gives an example
of the momentum distribution and the spin polarization of a trapped
spin-orbit-coupled BEC as a function of the Raman coupling. For
simplicity, we have considered harmonic trapping only along the
$x$ direction. One can see that the three phases discussed in the
bulk case show up also here.
It is worth mentioning that in the low density limit, where the
interaction energy is much smaller than the recoil energy, the value
of $\Omega/k_0^2$ at the transitions (\ref{eq:OmegaI-II}) and
(\ref{eq:OmegaII-III}) is almost density-independent, therefore even
in the presence of a trap they can be well identified using the
results obtained in the bulk.

\subsection{Dipole mode: a sum-rule approach}
\label{subsec:Dipole_sumrule}

Among the various excitations exhibited by a trapped
spin-orbit-coupled gas, the dipole mode deserves a special attention.
It corresponds to the oscillation of the center-of-mass of
the system, and can be easily excited experimentally
\cite{Kurn1998}. For a conventional trapped gas without spin-orbit
coupling, the oscillation along a certain direction, for example the
$x$ axis, is excited by the dipole operator $X= \sum_j x_j$,
and its frequency is equal to the frequency $\omega_x$ of the harmonic
trap. In the presence of spin-orbit coupling, the behavior of the
dipole oscillation can be studied using the formalism of sum
rules \cite{Li2012EPL}. A major advantage of this method is that
it can reduce the calculation of the dynamical properties of the
many-body system to the knowledge of a few key parameters relative
to the ground state.

The starting point of our analysis is represented by the definition
of the $k$-th moment of the dynamic structure factor for a general
operator $F$, given at zero temperature by\footnote{At finite
temperature, the moments $m_k$ should include the proper
Boltzmann factors \cite{bookPitaevskii2003}.}
\begin{equation}
m_k(F)= \sum_n (E_n-E_0)^k\,|\langle0|F|n\rangle|^2 .
\label{eq:moment_k}
\end{equation}
Here $|0\rangle$ and $|n\rangle$ are, respectively, the ground
state and the $n$-th excited state of the many-body Hamiltonian
(\ref{eq:H_N_body}), now including the external trapping potential
in the single-particle contributions
\begin{equation}
h_0^{\rm SO}(j)= \frac{1}{2}\left[\left(p_{x,j}-k_0 \sigma_{z,j}
\right)^2 + p_{\perp,j}^2 \right] + \frac{\Omega}{2} \sigma_{x,j}
+ \frac{\delta}{2} \sigma_{z,j}
+ V_{\rm ext}({\bf r}_j) ,
\label{eq:h_0_SO_trap}
\end{equation}
and $E_0$, $E_n$ are the corresponding energies. The quantity
$|\langle 0 |F| n \rangle|^2$ is called the strength of the
operator $F$ relative to the state $|n\rangle$.

Some moments can be easily calculated by employing the closure
relation and the commutation rules involving the Hamiltonian
of the system. In the case of the dipole operator $F=X$ one
finds, for example, that the energy-weighted moment takes the
well-known model-independent value (also called $f$-sum rule)
\begin{equation}
m_1(X)= \frac{1}{2} \langle0|\left[X,\left[H,X\right]\, \right]|0
\rangle = \frac{N}{2}
\label{eq:m_p1_X}
\end{equation}
with $N$ the total number of atoms. Notice that this sum rule is
not affected by the spin terms in the Hamiltonian, despite the
fact that the commutator of $H$ with $X$ explicitly depends on
the spin-orbit coupling:
\begin{equation}
\left[H,\,X\right] = -i\left(P_x-k_0 \Sigma_z\right) ,
\label{eq:comm_H_X}
\end{equation}
where $P_x= \sum_j p_{x,j}$ is the total momentum of the gas along
the $x$ direction, and $\Sigma_z = \sum_j\sigma_{z,j}$ is the
total spin operator along $z$. Equation (\ref{eq:comm_H_X})
actually reflects the fact that the equation of continuity
(and hence in our case the dynamic behavior of the center-of-mass
coordinate) is deeply influenced by the coupling with the spin
variable.

Another important sum rule is the inverse energy-weighted sum rule
(also called dipole polarizability). In the presence of harmonic
trapping, this sum rule can be calculated in a straightforward way
using the commutation relation
\begin{equation}
\left[H,\,P_x\right]=i\omega^2_x X
\label{eq:comm_H_Px}
\end{equation}
and the closure relation. One finds
\begin{equation}
m_{-1}(X)= \frac{m_1(P_x)}{\omega_x^4}= \frac{N}{2 \omega^2_x} .
\label{eq:m_m1_X}
\end{equation}
Both sum rules (\ref{eq:m_p1_X}) and (\ref{eq:m_m1_X}) are
insensitive to the presence of the spin terms in the single-particle
Hamiltonian (\ref{eq:h_0_SO_trap}), as well as to the
two-body interaction. This does not mean, however, that the
dipole dynamics is not affected by the spin-orbit
coupling. This effect is accounted for by another sum rule,
particularly sensitive to the low-energy region of the excitation
spectrum: the inverse cubic energy-weighted sum rule, for which
we find the exact result
\begin{equation}
m_{-3}(X)= \frac{m_{-1}(P_x)}{\omega_x^4} =\frac{N}{2
\omega^2_x} \left( 1+ k_0^2\,\chi^{}_M \right) ,
\label{eq:m_m3_X}
\end{equation}
where $\chi^{}_M$ corresponds to the magnetic polarizability already
defined in Sec.~\ref{subsec:Polar_compress}, and given in terms of
sum rules by $\chi^{}_M = 2 m_{-1}(\Sigma_z)/N$.
It is worth mentioning that the above results
for the sum rules $m_1(X)$, $m_{-1}(X)$ and $m_{-3}(X)$
hold exactly for the Hamiltonian (\ref{eq:H_N_body}), including the
interaction terms. Their validity is not restricted to the
mean-field approximation and is ensured for both Bose and Fermi
statistics, at zero as well as at finite temperature. In particular
the sum rule $m_{-3}(X)$, being sensitive to the magnetic
polarizability, is expected to exhibit a nontrivial temperature
dependence across the BEC transition.

\begin{figure}[t]
\centering
\includegraphics[scale=1]{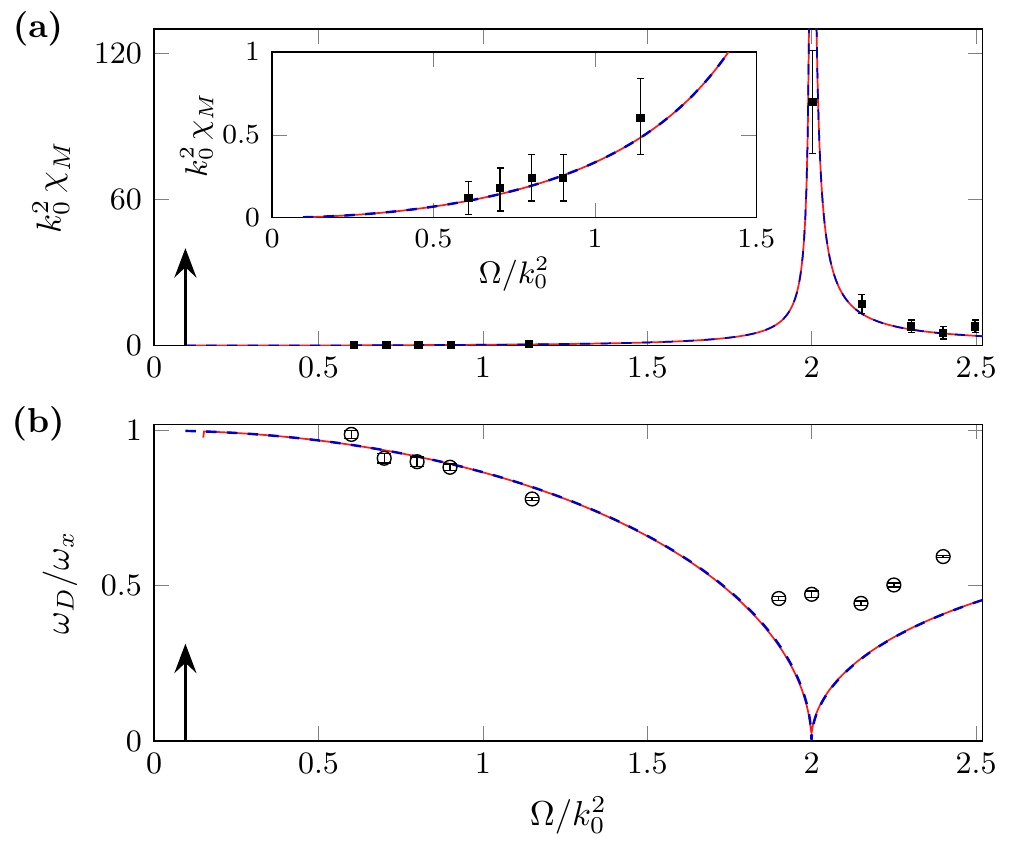}
\caption{{\bf (a)}. Magnetic polarizability $\chi^{}_M$ as a function
of $\Omega$ calculated in a trap (red solid lines) and in uniform
matter using the density in the center of the trap (blue dashed lines).
{\bf (b)}. Dipole frequency predicted by (\ref{eq:omega_D2}), using
the values of $\chi^{}_M$ shown above, represented by the red solid
lines and the blue dashed lines respectively. The parameters are
$k_0^2 = 2\pi\times 4.42\,$kHz, $\omega_x = 2\pi\times 45\, $Hz, the
scattering lengths $a_{\uparrow \uparrow} =a_{\downarrow \downarrow}
=101.20\, a_B$, $a_{\uparrow \downarrow}= 100.94\, a_B$, where $a_B$
is the Bohr radius, and the atomic mass of $^{87}$Rb. The density in the
center of the trap is $n \simeq 1.37\times 10^{14}\, $cm$^{-3}$. The
black squares and circles in the figures are the experimental data
of Ref.~\citen{Zhang2012}. The black arrows indicate the transition
between phases I and II.} \label{fig:Chen_exp}
\end{figure}

Equation (\ref{eq:m_m3_X}) exploits the crucial role played by the
spin-orbit coupling proportional to $k_0$. The effect is
particularly important when the magnetic polarizability takes a large
value. A large increase of $\chi^{}_M$ is associated with the
occurrence of a dipole soft mode as can be inferred by taking the
ratio between the inverse and cubic inverse energy-weighted sum
rules $m_{-1}(X)$ and $m_{-3}(X)$, yielding the rigorous upper
bound
\begin{equation}
\omega^{2}_D = \frac{m_{-1}(X)}{m_{-3}(X)} =
\frac{\omega_x^2}{1 + k_0^2\,\chi^{}_M}
\label{eq:omega_D2}
\end{equation}
to the lowest dipole excitation energy. The value of $\chi^{}_M$ for a
trapped BEC can be calculated in the same way as in uniform matter
(see Sec.~\ref{subsec:Polar_compress}), with the difference that
the condensate wave function is now provided by the solution of
Eq.~(\ref{eq:GPE}) rather than by the ansatz (\ref{eq:ansatz}). Figure
\ref{fig:Chen_exp}(a) shows the behavior of the magnetic polarizability
in the plane-wave and the single-minimum phases as a function of the
Raman coupling $\Omega$, calculated by numerically solving
Eq.~(\ref{eq:GPE}) in the presence of harmonic trapping along the $x$
direction (red dashed lines), and by the relations
(\ref{eq:chi_M_II}) and (\ref{eq:chi_M_III}) in uniform matter
using the density in the center of the trap (blue solid lines).
The choice of the parameters corresponds to the
experimental conditions of Ref.~\citen{Zhang2012}. The black squares
represent the magnetic polarizability extracted from the
measurement of the oscillation amplitudes of some relevant
quantities \cite{Zhang2012}
(see the discussion in Sec.~\ref{subsec:oscill_ampl}).
Figure \ref{fig:Chen_exp}(b) shows the frequency
of the dipole oscillation predicted from Eq.~(\ref{eq:omega_D2})
using the same values of $\chi^{}_M$ presented in (a). It reveals
important deviations from the trap frequency $\omega_x$ caused
by the spin-orbit coupling. The circles are the experimental results
of Ref.~\citen{Zhang2012}. Far from the transition point at $\Omega \simeq
2k_0^2$ the theoretical curves agree very well with the
experimental data, while near the transition nonlinear effects
play a major role, as discussed in Ref.~\citen{Zhang2012}. The lack of
data points in the region below the transition is due to the
occurrence of a dynamic instability, which makes the observation
of the dipole oscillation very difficult \cite{Ozawa2013}.

\subsection{Dipole mode and oscillation amplitudes}
\label{subsec:oscill_ampl}

The combined spin-orbit nature of the lowest dipole mode is also
nicely revealed by the relative amplitudes of the oscillating values
of the center-of-mass position ($A_X$),
the momentum ($A_{P_x}$) and the spin polarization ($A_{\Sigma_z}$).
These amplitudes can be calculated in the present
approach by writing the many-body oscillating wave function as
\begin{equation}
|\Psi(t)\rangle= e^{i\alpha(t) \delta P_x}e^{\beta(t)G} |0\rangle ,
\label{eq:psi_dynamic_polar}
\end{equation}
where $\delta P_x = P_x - \left\langle P_x\right\rangle_0$ plays the role
of the excitation operator, $\left\langle \; \right\rangle_0$ denoting
the expectation value on the ground state $\left|0\right\rangle$,
while $G$ represents the restoring force defined by the commutation relation
$[H,\,G]= \delta P_x$, and $\alpha, \, \beta$ are time-dependent
parameters. The equations governing the time evolution of these
parameters can be obtained through a variational Lagrange procedure;
at the lowest order in $\alpha$ and $\beta$ they read
\begin{align}
\dot{\alpha}(t) &= -\beta(t) , \label{eq:amplitude_alpha}\\
\dot{\beta}(t) &= \frac{\omega^2_x}{1+ k_0^2\, \chi^{}_M} \, \alpha(t) .
\label{eq:amplitude_beta}
\end{align}
The time dependence of the relevant quantities
$\langle X \rangle$, $\langle P_x \rangle$ and $\langle \Sigma_z
\rangle$ can be obtained by solving Eqs.~(\ref{eq:amplitude_alpha})
and (\ref{eq:amplitude_beta}). The relations between the spin, the
center-of-mass and the momentum oscillation amplitudes eventually
take the useful form
\begin{align}
A_{\Sigma_z} &= A_X \frac{\omega_x \, k_0\, \chi^{}_M}
{\sqrt{1+k_0^2\,\chi^{}_M}} \, ,\label{eq:amplitude_1}\\
\frac{A_{P_x}}{k_0} &= A_{\Sigma_z} \frac{1 +k_0^2\, \chi^{}_M}
{k_0^2\,\chi^{}_M} \, .\label{eq:amplitude_2}
\end{align}
The connection between the momentum and spin amplitudes has been
already pointed out in Ref.~\citen{Zhang2012} (see Fig.~4 therein).
It provides a practical way to
determine experimentally the magnetic polarizability $\chi^{}_M$.
Near the transition point between the plane-wave and the single-minimum
phase the ratio $A_{\Sigma_z}/A_X$ between the spin and the
center-of-mass amplitudes diverges like $\sqrt{\chi^{}_M}$, in analogy
with the behavior exhibited by the ratio between the spin and the
density amplitudes in the propagation of sound
(see Eq.~(\ref{eq:amplitude_trans})).

It is worth pointing out that the results
(\ref{eq:amplitude_1})--(\ref{eq:amplitude_2}), as well as the upper
bound of the excitation frequency (\ref{eq:omega_D2}), are expected to
be accurate when the Raman coupling $\Omega$ is larger than the
trapping frequency $\omega_x$.
Instead, in the opposite limit $\Omega \ll \omega_x$,
the lowest mode is mainly a spin oscillation,
which does not exhibit any significant coupling to the center-of-mass
motion. The corresponding frequency can be estimated with
a sum-rule approach by considering the ratio $m_{1}(\Sigma_z) /
m_{-1}(\Sigma_z)$ of the moments of the spin operator $\Sigma_z$.
The excitation frequency calculated in this way is found to
vanish linearly with $\Omega$. Finally, in the intermediate regime
between the two limits discussed above, one can define
$F=P_x+\eta k_0 \Sigma_z$ and use the ansatz $\delta F =
F - \left\langle F \right\rangle_0$ for the operator exciting the dipole
oscillation, where the value of the variational parameter $\eta$
is found by minimizing the estimate $m_{1}(F) / m_{-1}(F)$
for the excitation frequency. The corresponding oscillation amplitudes of
the relevant physical quantities can be calculated by a procedure
analogous to the one discussed above \cite{Li2012EPL}.

\subsection{Hydrodynamic formalism}
\label{subsec:Hydro_form}

A useful approach to describe the phonon regime in the excitation
spectrum of a superfluid is provided by hydrodynamic theory.
For a spinor BEC this theory can be derived by writing
the spin-up and spin-down components of the order parameter in
terms of their modulus and phase \cite{Zheng2012,Martone2012}.
In the resulting equations the quantum pressure terms can be
safely neglected in the phonon regime, characterized by long
wavelengths and low frequencies. Furthermore, since the phonon
frequencies are much smaller than the gap between the two branches
of the excitation spectrum, which is of the order of $\Omega$,
the relative phase of the two spin components is locked
($\phi_{\uparrow}=\phi_{\downarrow}$). As a consequence, the
relevant hydrodynamic equations reduce to the equations for the change
in the total density $\delta n$ and in the phase $\delta\phi =
\delta\phi_{\uparrow} = \delta\phi_{\downarrow}$. Assuming for
simplicity $g_{\uparrow\downarrow}=g$, these two equations assume
the simple form
\begin{equation}
\frac{\partial}{\partial t}\delta n + \nabla_{\perp} \cdot \left(n
\,\nabla_{\perp} \delta\phi \right) + \frac{m\;}{m^\ast} \partial_x
\left( n \,\partial_x \delta\phi \right)=0
\label{eq:HD_continuity}
\end{equation}
and
\begin{equation}
\frac{\partial}{\partial t}\delta\phi + \delta\mu = 0 ,
\label{eq:HD_Eular}
\end{equation}
with $m/m^\ast$ given by Eqs.~(\ref{eq:mass_12}) and (\ref{eq:mass_3})
and $\delta\mu = g\delta n$. Notice that, due to the assumption
$g_{\uparrow\downarrow}=g$, the above hydrodynamic picture can
describe the dynamics only in the plane-wave and the single-minimum
phases (the investigation of the phonon modes in the stripe phase
requires a more
sophisticated calculation, see Sec.~\ref{sec:Excitation_stripe}).
Remarkably, the equation of continuity (\ref{eq:HD_continuity}) is
crucially affected by the spin-orbit coupling. This follows from
the fact that the current is not simply given by the canonical
momentum operator, as happens in usual superfluids, but
contains an additional spin contribution, accounted for, in
Eq.~(\ref{eq:HD_continuity}), through the effective mass term.
The current density operator should actually satisfy the continuity
equation $[H,\,\rho({\bf r})]=i\nabla\cdot{\bf j}$, where
$\rho({\bf r})=\sum_j\delta({\bf r}-{\bf r}_j)$ is the
total density operator. By explicitly carrying out the commutator
one identifies the current as ${\bf j}({\bf r})={\bf p}({\bf r})
-k_0\sigma_z({\bf r})\hat{\bf e}_x$, where ${\bf p}({\bf r})=\sum_j
[{\bf p}_j \, \delta({\bf r}-{\bf r}_j)+{\rm H.c.}]/2$ and
$\sigma_z({\bf r})=\sum_j\sigma_{z,j}\,\delta({\bf r}-{\bf r}_j)$
are the momentum and spin-density operators, respectively. This
expression for the current explicitly reveals the presence of a
gauge field associated to the vector potential ${\bf A}=
k_0\sigma_z \hat{\bf e}_x\,$. It is worth noticing that at
equilibrium the momentum and spin-dependent terms exactly
compensate each other, yielding $\langle{\bf j}({\bf r})\rangle=0$.
The presence of the spin term in the current also reflects
the violation of Galilean invariance in the spin-orbit Hamiltonian
\cite{Ozawa2013}.

Combining (\ref{eq:HD_continuity}) and (\ref{eq:HD_Eular}) one finds
the following equation for the density:
\begin{equation}
\frac{\partial^2}{\partial t^2}\delta n = g\left[\nabla_{\perp} \cdot
\left(n\, \nabla_{\perp} \delta n \right)+ \frac{m\;}{m^\ast}
\partial_x \left(n\,\partial_x \delta n\right) \right] .
\label{eq:HD}
\end{equation}
In uniform matter, characterized by a constant density $n = \bar{n}$,
Eq.~(\ref{eq:HD}) yields the relation $c^2 =
g\bar{n}/m^\ast$ for the sound velocity along the $x$ direction,
consistent with the results (\ref{eq:sound_velo_II}) and
(\ref{eq:sound_velo_III}) for $g_{\uparrow\downarrow}=g$. In the
presence of harmonic trapping, where the equilibrium density profile
is given by an inverted parabola, the solutions of the hydrodynamic
equation (\ref{eq:HD}) coincide with those one finds for usual
BECs, with the simple replacement of the trap frequency $\omega_x$
with $\omega_x\sqrt{m/m^\ast} \,$. This gives the result
$\omega_D = \omega_x \sqrt{m/m^\ast} \,$ for the dipole frequency,
which is consistent with the estimate (\ref{eq:omega_D2}) based on
a sum-rule approach, once the relation (\ref{eq:chi_mass}) between
the effective mass and the magnetic polarizability (holding for
$G_2=0$) is taken into account. Equation (\ref{eq:HD}) also shows that,
for any other hydrodynamic mode involving a motion of the gas along
the $x$ axis, a similar effect of strong reduction of the frequency
close to the second-order transition should be expected. This is the case,
for example, of the scissors mode for deformed traps in the $x$-$y$ or
$x$-$z$ plane, where the collective frequency takes the form
$\sqrt{(m/m^\ast)\omega_x^2+\omega_y^2}$ and
$\sqrt{(m/m^\ast)\omega_x^2+\omega_z^2}$ respectively.

\section{Static and Dynamic Properties of the Stripe Phase}
\label{sec:Excitation_stripe}

The stripe phase is doubtlessly the most intriguing phase appearing
in the phase diagram of Sec.~\ref{sec:Ground_state}. It has been
the object of several recent theoretical investigations
\cite{Wang2010,Ho2011,Wu2011,Sinha2011,Ozawa2012,Li2013,
Zezyulin2013,Lan2014,Sun2014,Han2014,Hickey2014}.
As we already pointed out,
the stripe phase is characterized by the spontaneous breaking of
two continuous symmetries. The breaking of gauge symmetry yields
superfluidity, while the breaking of translational invariance is
responsible for the occurrence of a crystalline structure. The
simultaneous presence of these two broken symmetries is typical of
supersolids
\cite{Boninsegni2012,Andreev1969,Leggett1970,Chester1970}.
As we shall see, it is at the origin of the appearance of two
gapless excitations as well as of a band structure in the
excitation spectrum \cite{Li2013}.

Some important properties of the ground state and the dynamics
of the stripe phase in uniform matter will be discussed in
Secs.~\ref{subsec:Bogoliubov_stripe} and \ref{subsec:Structure_stripe}.
Many relevant quantities that we will consider, such as the contrast of
the density modulations (\ref{eq:contrast}), will turn out to depend
crucially on the value of the Raman coupling $\Omega$. Therefore,
in order to enhance the effects of the presence of the stripes one
needs to use relatively large values of $\Omega$. On the other hand,
the stripe phase is favored only in a range of low values of the Raman
coupling lying below the transition frequency $\Omega^{({\rm I-II})}$.
In the following we will consider configurations with relatively large
values of the parameter $G_2$ which, as can be seen from
Eq.~(\ref{eq:OmegaI-II}), allow to obtain a significant increase of
the critical value of $\Omega$. This is not, however, the situation in
current experiments with $^{87}$Rb atoms \cite{Lin2011,Zhang2012},
where $G_2$ is instead extremely small. In Sec.~\ref{subsec:Exp_stripes}
we will illustrate a procedure to increase the value of $G_2$ with
available experimental techniques.

\subsection{Ground state and excitation spectrum}
\label{subsec:Bogoliubov_stripe}

\begin{figure}[t]
\centering
\includegraphics[scale=1]{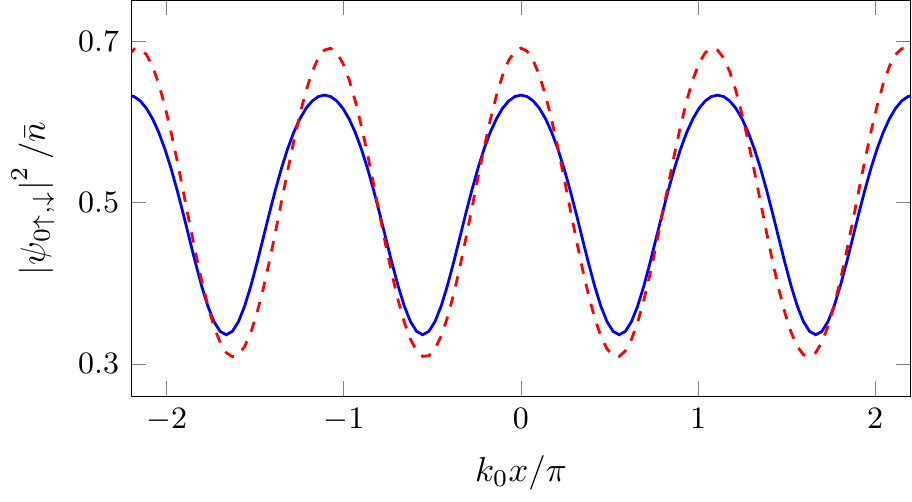}
\caption{Density profile in the stripe phase along the $x$
direction, calculated within the first-order harmonic approximation
(\ref{eq:ansatz}) (red dashed line) and from Eq.~(\ref{eq:psi0_stripe})
including the higher-order harmonics (blue solid line). The
parameters are $\Omega/k_0^2=1.0$, $G_1/k_0^2=0.3$, and $ G_2/k_0^2=
0.08$, yielding the transition frequency $\Omega^{({\rm I-II})}
/k_0^2 \simeq 1.3$.} \label{fig:dens_stripe}
\end{figure}

In Sec.~\ref{subsec:Many_body} the ground state in the
stripe phase has been described by means of an approximated wave
function, based on the ansatz (\ref{eq:ansatz}), which takes into
account only first-order harmonic terms. The exact wave function
includes also higher-order harmonics, whose appearance is a
consequence of the nonlinearity of the Gross--Pitaevskii theory. It
can be written in the form
\begin{equation}
\begin{pmatrix} \psi_{0\uparrow}\\ \psi_{0\downarrow} \end{pmatrix}
= \sqrt{\bar{n}}
\sum_{\bar{K}} \begin{pmatrix} \;\;\, a_{-k_1+\bar{K}} \, \\
-b_{-k_1+\bar{K}}\, \end{pmatrix} e^{i \left(\bar{K} - k_1 \right)
x} ,
\label{eq:psi0_stripe}
\end{equation}
where $k_1 = \pi/d$ is related to the period $d$ of the stripes,
$\bar{K}=2 n k_1$, with $n=0, \,\pm 1, \,\ldots\,$, are the
reciprocal lattice vectors, while $a_{-k_1+\bar{K}}$ and
$b_{-k_1+\bar{K}}$ are expansion coefficients to be determined,
together with the value of $k_1$, by a procedure of minimization
of the mean-field energy functional (\ref{eq:E_tot}). The
energy minimization gives rise to the presence of terms with
opposite phase ($e^{\pm ik_1x}, \,e^{\pm 3ik_1x},\, \ldots\,$),
responsible for the density modulations and characterized by the
symmetry condition $a_{-k_1+\bar{K}}=b^\ast_{k_1-\bar{K}}$,
causing the vanishing of the spin polarization $\langle\sigma_z\rangle$.
Figure \ref{fig:dens_stripe} shows an example of density profile in
the stripe phase, calculated for a configuration with relatively large
values of $G_2$ and $\Omega/k_0^2$ in order to emphasize the contrast
in the density modulations.

As in the case of the uniform phases, also in the stripe phase
we can evaluate the elementary excitations by the standard
Bogoliubov approach, writing the deviations of the order
parameter with respect to equilibrium as
\begin{equation}
\Psi= e^{-i\mu t}\left[\begin{pmatrix} \psi_{0\uparrow} \\
\psi_{0\downarrow}\end{pmatrix} + \begin{pmatrix}
u_{\uparrow}({\bf r}) \\ u_{\downarrow}({\bf r})
\end{pmatrix} e^{-i\omega t} + \begin{pmatrix} v^\ast_{\uparrow}
({\bf r})\\ v^\ast_{\downarrow}({\bf r}) \end{pmatrix}
e^{i\omega t}\right]
\label{eq:spinor_stripe}
\end{equation}
and solving the corresponding linearized time-dependent
Gross--Pitaevskii equations. The equations are conveniently
solved by expanding $u_{\uparrow, \downarrow}({\bf r})$ and
$v_{\uparrow, \downarrow}({\bf r})$ in the Bloch form in terms
of the reciprocal lattice vectors:
\begin{eqnarray}
u_{{\bf q}\,\uparrow,\downarrow}({\bf r}) &=&
e^{-ik_1 x} \sum_{\bar{K}}
U_{{\bf q}\,\uparrow,\downarrow\,\bar{K}}
\,e^{i{\bf q}\cdot {\bf r}+i\bar{K} x} ,\\
v_{{\bf q}\,\uparrow,\downarrow}({\bf r}) &=&
e^{ik_1 x} \sum_{\bar{K}}
V_{{\bf q}\,\uparrow,\downarrow\,\bar{K}} \,e^{i{\bf q}\cdot
{\bf r}-i\bar{K} x} ,
\end{eqnarray}
where ${\bf q}$ is the wave vector of the excitation. This
ansatz can also be used to calculate the density and
spin-density dynamic response function, similarly to what
we did in Sec.~\ref{subsec:Bogoliubov_uniform},
by adding to the Hamiltonian a perturbation proportional to
$e^{i({\bf q} \cdot
{\bf r}-\omega t) +\eta t}$ and $\sigma_z e^{i({\bf q} \cdot
{\bf r}-\omega t) +\eta t}$ with $\eta \to 0^+$, respectively.

\begin{figure}
\centering
\includegraphics[scale=1]{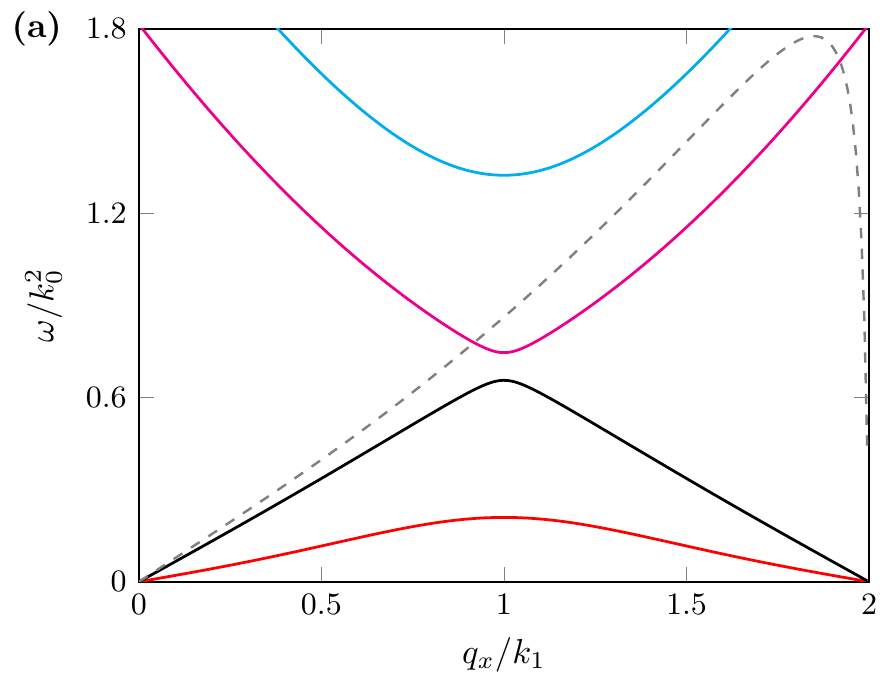}
\includegraphics[scale=1]{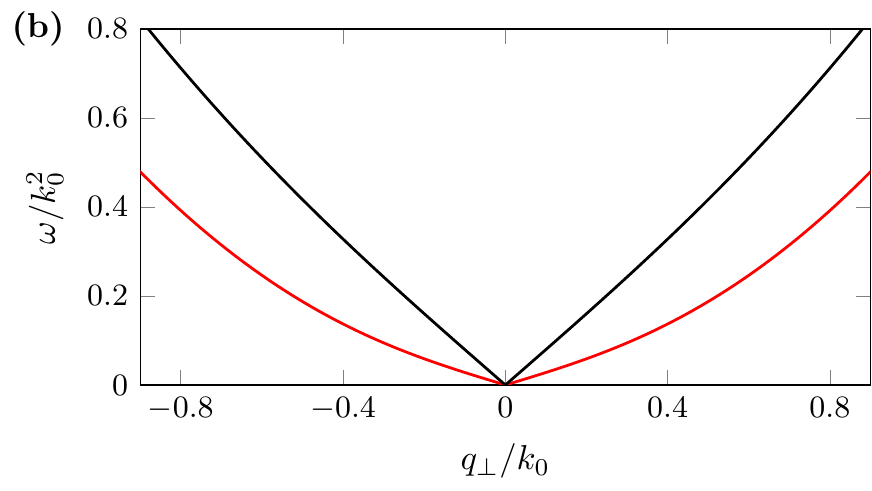}
\caption{{\bf (a)}. Lowest four excitation bands (solid lines) along
the $x$ direction ($q_{\perp}=0$). The dashed line corresponds to
the Feynman relation $\omega = q_x^2/2 S(q_x)$. {\bf (b)}. Lowest
two excitation bands in the transverse direction ($q_x =0$). The
parameters are the same as in Fig.~\ref{fig:dens_stripe}.}
\label{fig:spec_stripe}
\end{figure}

The spectrum of the elementary excitations in the stripe phase is
reported in Fig.~\ref{fig:spec_stripe} for the same parameters
used in Fig.~\ref{fig:dens_stripe}. We have considered both
excitations propagating in the $x$ direction orthogonal to the
stripes (labelled with the wave vector $q_x$) and in the
transverse directions parallel to the stripes (identified by the
wave vector $q_{\perp}$). A peculiar feature, distinguishing the
stripe phase from the other uniform phases, is the occurrence of two
gapless bands. The excitation energies along the
$x$ direction vanish at the Brillouin wave vector $q_B=2k_1$,
which is a usual situation in crystals. A similar double gapless band
structure has been predicted recently in condensates with soft-core,
finite-range interactions \cite{Saccani2012,Macri2013,Kunimi2012}.

\begin{figure}[t]
\centering
\includegraphics[scale=1]{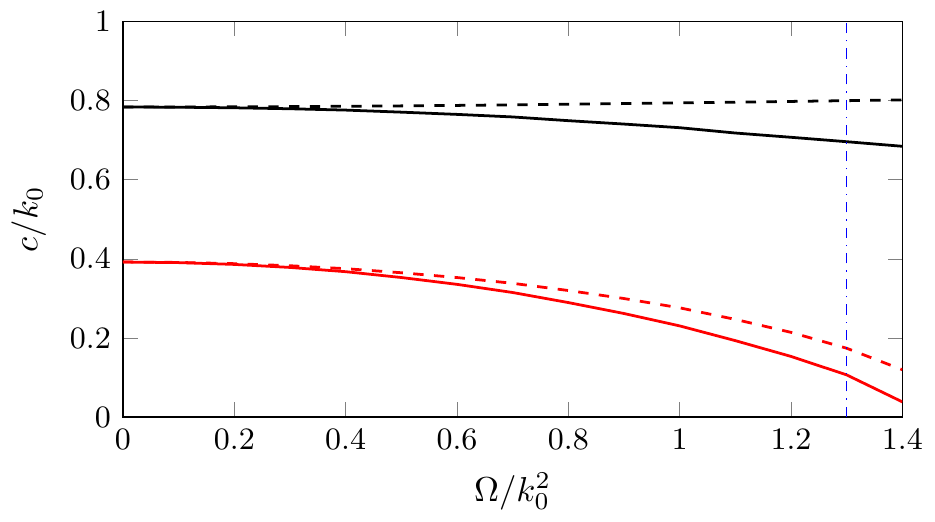}
\caption{Sound velocities in the first (red) and second (black)
bands along the $x$ ($c_x$, solid lines) and transverse
($c_{\perp}$, dashed lines) directions as a function of $\Omega$.
The blue dash-dotted line represents the transition from the stripe
phase to the plane-wave phase. The values of the parameters
$G_1/k^2_0$ and $G_2/k^2_0$ are the same as in
Fig.~\ref{fig:dens_stripe}.} \label{fig:sound_velo_stripe}
\end{figure}

In Fig.~\ref{fig:sound_velo_stripe} we compare the sound velocities
of the two gapless branches in the longitudinal ($c_x$) and
transverse ($c_\perp$) directions. We find that $c_x$ is always
smaller than $c_\perp$, reflecting the inertia of the flow caused by
the presence of the stripes. The value of $c_\perp$ in the second
band (second sound) is well reproduced by the Bogoliubov expression
$\sqrt{2G_1}$ (equal to $0.78 \,k_0$ in our case) for the sound
velocity. Notice that the sound velocity in the first band (first
sound) becomes lower and lower as the Rabi frequency increases,
approaching the transition to the plane-wave phase. The Bogoliubov
solutions in the stripe phase exist also for values of $\Omega$
larger than the critical value $\Omega^{({\rm I-II})}= 1.3
\,k_0^2$, due to the first-order nature of the transition
(effect of metastability).

\subsection{Static structure factor and static response function}
\label{subsec:Structure_stripe}

\begin{figure}[t]
\centering
\includegraphics[scale=1]{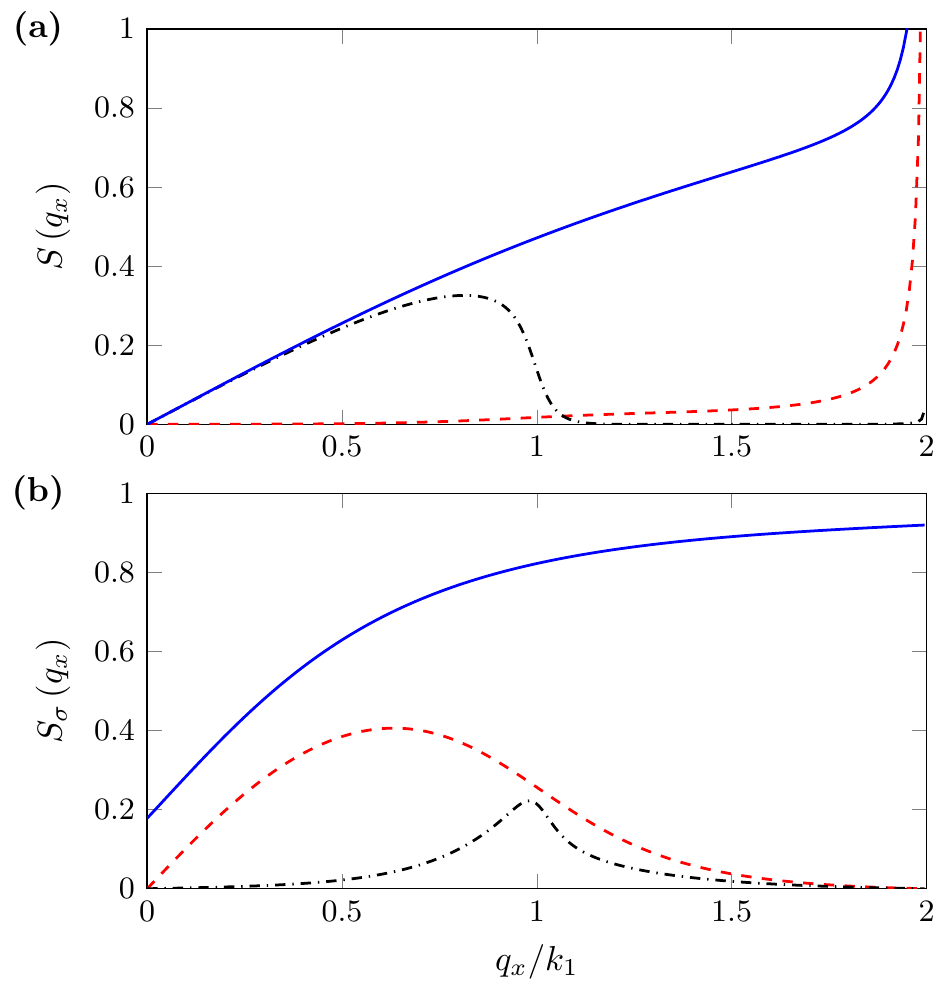}
\caption{Density {\bf (a)} and spin-density {\bf (b)} static
structure factor as a function of $q_x$ (blue solid line). The
contributions of the first (red dashed line) and second (black
dash-dotted line) bands are also shown. The parameters are the same
as in Fig.~\ref{fig:dens_stripe}.} \label{fig:static_struc}
\end{figure}

The nature of the excitation bands can be understood by
calculating the static structure factors for the density and
the spin-density operators, which can be written as
\begin{equation}
S({\bf q}) = N^{-1}\sum_\ell| \langle0|\rho_{\bf q}| \ell \rangle
|^2
\label{eq:S_dens}
\end{equation}
and
\begin{equation}
S_\sigma({\bf q})=N^{-1} \sum_\ell |\langle0|\sigma_{z,{\bf q}}|
\ell\rangle|^2
\label{eq:S_spin}
\end{equation}
respectively. In these equations $\rho_{\bf q}=\sum_j e^{i {\bf q}
\cdot {\bf r}_j}$ and $\sigma_{z,{\bf q}}=\sum_j \sigma_{z,j} e^{i
{\bf q} \cdot {\bf r}_j}$ are the ${\bf q}$ components of the
above-mentioned operators, while $\ell$ is the band index. In
Fig.~\ref{fig:static_struc} we show the static structure factors for
wave vectors along the $x$ axis, as well as the contributions to the
total sum coming from the two gapless branches ($\ell=1,\,2$). The
figure clearly shows that, at small $q_x$, the lower branch is
basically a spin excitation, while the upper branch is a density
mode. The density nature of the upper branch, at small $q_x$, is
further confirmed by the comparison with the Feynman relation
$\omega = q_x^2/2 S(q_x)$ (see Fig.~\ref{fig:spec_stripe}(a)). A
two-photon Bragg scattering experiment with laser frequencies far
from resonance, being sensitive to the density response, will
consequently excite only the upper branch at small $q_x$. Bragg
scattering experiments actually measure the imaginary part of the
response function, a quantity which, at enough low temperature, can
be identified with the $T=0$ value of the dynamic structure factor
$S(q_x,\omega)=\sum_\ell| \langle0|\rho_{q_x}| \ell\rangle|^2
\delta(\omega - \omega_{\ell 0})$, where $\omega_{\ell 0}$ is the
excitation frequency of the $\ell$-th state
\cite{bookPitaevskii2003}. Notice that, differently from $S(q_x)$,
the spin structure factor $S_{\sigma}(q_x)$ does not vanish as
$q_x\to 0$, being affected by the higher energy bands as a
consequence of the Raman term in Hamiltonian (\ref{eq:h_0_SO}). As
$q_x$ increases, the lower branch actually reveals a hybrid
character and, when approaching the Brillouin wave vector
$q_B=2k_1$, it is responsible for the divergent behavior of the
density static structure factor (see Fig.~\ref{fig:static_struc}(a)),
which is again a typical feature exhibited by crystals.

\begin{figure}[t]
\centering
\includegraphics[scale=1]{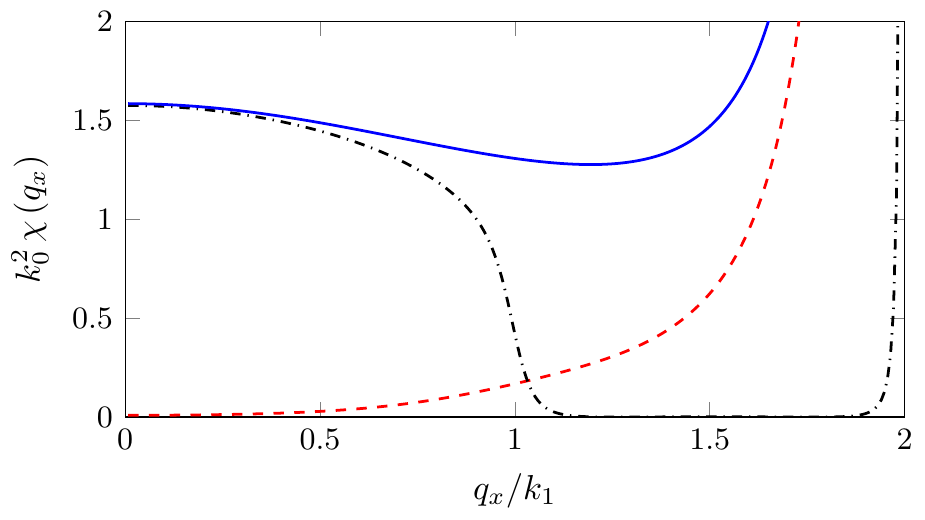}
\caption{Static response as a function of $q_x$ (blue solid line).
The contributions of the first (red dashed line) and second (black
dash-dotted line) bands are also shown. The parameters are the same
as in Fig.~\ref{fig:dens_stripe}.} \label{fig:dens_chi}
\end{figure}

It is worth pointing out that the occurrence of two gapless
excitations is not by itself a signature of supersolidity and is
exhibited also by uniform mixtures of BECs without spin-orbit and
Raman couplings \cite{bookPethick2008} as well as by the plane-wave
phase of the Rashba Hamiltonian with $SU(2)$-invariant interactions
($G_2=0$) \cite{Barnett2012,Xu2012,Liao2013}. Only the occurrence of
a band structure, characterized by the vanishing of the excitation
energy and by the divergent behavior of the structure factor at the
Brillouin wave vector, can be considered an unambiguous evidence of
the density modulations characterizing the stripe phase. The
divergent behavior near the Brillouin zone is even more pronounced
(see Fig.~\ref{fig:dens_chi}) if one investigates the static
response function
\begin{equation}\chi(q_x) = 2 N^{-1} \sum_\ell
\frac{|\langle0| \rho_{q_x}|\ell\rangle|^2}{\omega_{\ell 0}} \, ,
\label{eq:chi_stat_stripe}
\end{equation}
proportional to the inverse energy-weighted moment of the dynamic
structure factor.

The divergent behaviors of $S(q_x)$ and $\chi(q_x)$ can be
rigorously proven using the Bogoliubov \cite{Bogoliubov1962} and the
uncertainty principle \cite{Pitaevskii1991,Pitaevskii1993}
inequalities applied to systems with spontaneously broken continuous
symmetries. These inequalities are based, respectively, on the
relations
\begin{equation}
m_{-1}(F) m_1(G) \ge |\langle\,\left[F,\, G\right]\,
\rangle|^2
\label{eq:Bogoliubov}
\end{equation}
and
\begin{equation}
m_0(F) m_0(G) \ge |\langle\, \left[F,\,G\right]\, \rangle|^2
\label{eq:uncertainty}
\end{equation}
involving the $k$-th moments $m_k(\mathcal{O})=\sum_\ell
\left(|\langle0|\mathcal{O}| \ell\rangle|^2+|\langle
0|\mathcal{O}^\dag|\ell\rangle|^2 \right)\omega_{\ell 0}^k$ of the
$\ell$-th strengths of the operators $F=\sum_j \,e^{i q_x x_j}$ and
$G = \sum_j(p_{x,\,j}\, e^{-i(q_x -q_B)x_j} + {\rm H.c.})/2$, with
$q_B=2k_1$ the Brillouin wave vector defined above. The commutator
$\langle\,[F,G]\,\rangle =q_x N\langle e^{i q_B x} \rangle$,
entering the right-hand side of the inequalities, coincides with the
relevant crystalline order parameter and is proportional to the
density modulations of the stripes. The moments $m_{-1}(F)$ and
$m_0(F)$ are instead proportional to the static response $\chi(q_x)$
and to the static structure factor $S(q_x)$, respectively. It is not
difficult to show that the moments $m_1(G)$ and $m_0(G)$ are
proportional, respectively, to $(q_x-q_B)^2$ and to $|q_x-q_B|$ as
$q_x \to q_B$ due to the translational invariance of the
Hamiltonian. This causes the divergent behaviors $S(q_x)\propto
1/|q_x-q_B|$ and $\chi(q_x)\propto 1/(q_x-q_B)^2$ with a weight
factor proportional to the square of the order parameter. The value
of the crystalline order parameter $\langle e^{i q_B x}\rangle$ is
larger for larger values of $\Omega$. For this reason it is useful
to work with large values of the spin interaction parameter $G_2$,
allowing for large values of the Raman coupling.\footnote{For
$^{87}$Rb the value of $G_2$ is small and the divergency effect in
$S(q_x)$ is weak. In this case, the sound velocity of the lowest
band is small and the dispersion practically exhibits a $q^2$-like
behavior at small $q$.} The experimental achievement of
configurations with relatively large $G_2$ will be the subject of
the next subsection.

\subsection{Experimental perspectives for the stripe phase}
\label{subsec:Exp_stripes}

As we have already anticipated, there is still no experimental
evidence for the periodic modulations of the density profile in the
stripe phase. The main reason is that, in the conditions of current
experiments with spin-orbit-coupled $^{87}$Rb BECs \cite{Lin2011,
Zhang2012}, the contrast and the wavelength of the fringes are too
small to be revealed. Another problem originates from the smallness
of the difference $\Delta\mu$ between the chemical potentials in the
plane-wave and the stripe phases, which, assuming
$g_{\uparrow\uparrow} = g_{\downarrow\downarrow}$, is given by
$\Delta\mu=2G_2$ in the $\Omega = 0$ limit, and becomes even smaller
at finite $\Omega$. As a consequence, a tiny magnetic field
(arising, for example, from external fluctuations) can easily bring
the system into the spin-polarized plane-wave phase.

In Ref.~\citen{Martone2014a} we have proposed a procedure to make
the experimental detection of the fringes a realistic perspective,
improving their contrast and their wavelength, and increasing
the stability of the stripe phase against magnetic fluctuations.
The idea is to trap the atomic gas in a 2D configuration, with
tight confinement of the spin-up and spin-down components
around two different positions, displaced by a distance $d$ along
the $z$ direction. This configuration can be realized with a trapping
potential of the form
\begin{equation}
V_\mathrm{ext}(z) =
\frac{\omega^2_z}{2}\left(z-\frac{d}{2}\sigma_z\right)^2
\label{eq:Vextd}
\end{equation}
with a sufficiently large value of $\omega_z$. As a consequence
of these trapping conditions, the overlap of the densities of the two
spin components can be significantly quenched,
and thus the effective interspecies
coupling is reduced with respect to the intraspecies couplings.
This yields a value of the parameter $\gamma$ larger than in the
$d=0$ case, and consequently the critical Raman
coupling $\Omega^{({\rm I-II})}$ can significantly increase
(see Eq.~(\ref{eq:OmegaI-II})), allowing for the realization of
the striped configuration with a high fringe contrast
(\ref{eq:contrast}).\footnote{Another
important consequence is that, due to the increase of the
value of $\gamma$, the critical density $n^{(c)}$ can be
significantly lowered with respect to the value in the $d=0$
case, becoming of more realistic achievement in future experiments.}

\begin{figure}[t]
\centering
\includegraphics[scale=1.0]{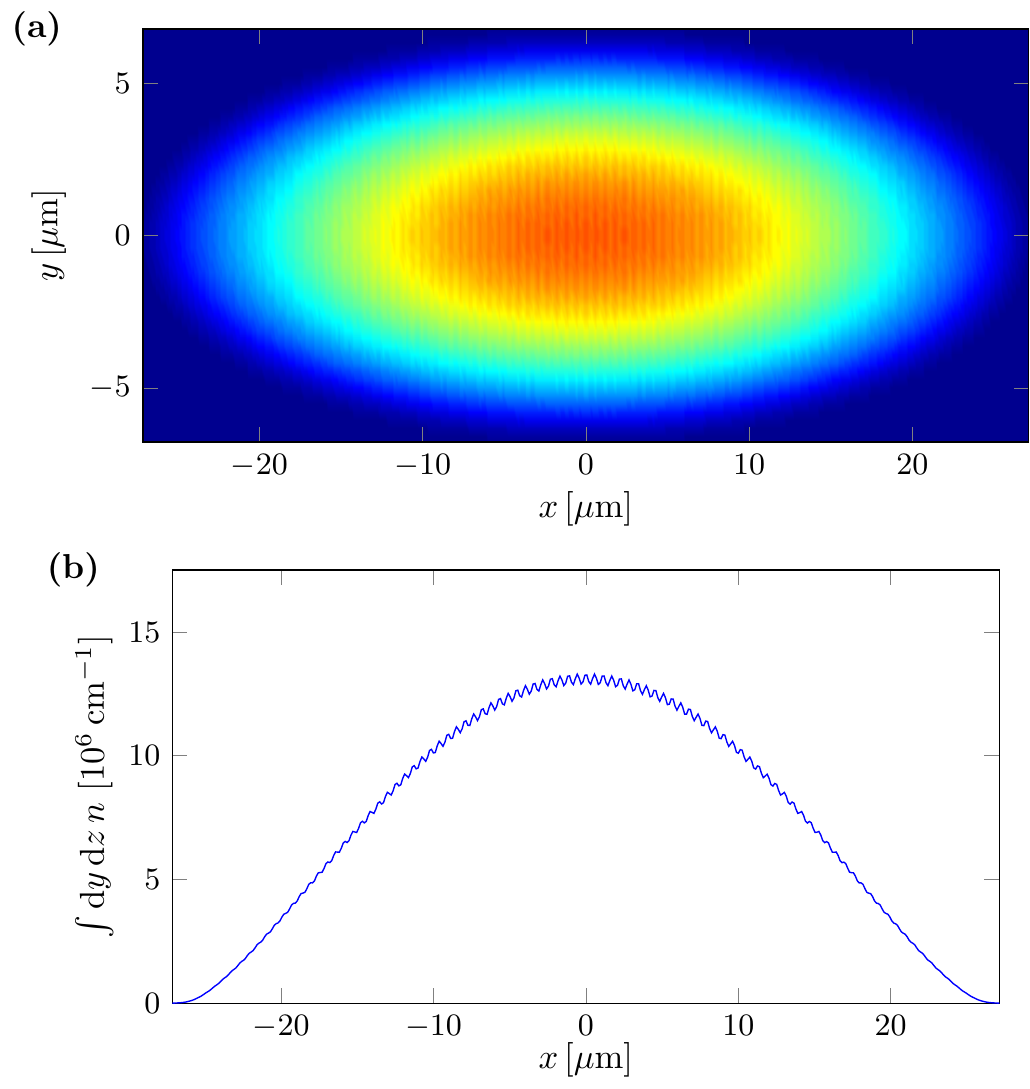}
\caption{Integrated density profiles $\int {\rm d}z \, n$ {\bf (a)}
and $\int {\rm d}y \, {\rm d}z \, n$ {\bf (b)} in the stripe phase,
evaluated in the conditions described in the text, and without
separation of the traps for the two spin components ($d=0$).}
\label{fig:density_prof_d0}
\end{figure}

\begin{figure}[t]
\centering
\includegraphics[scale=1.0]{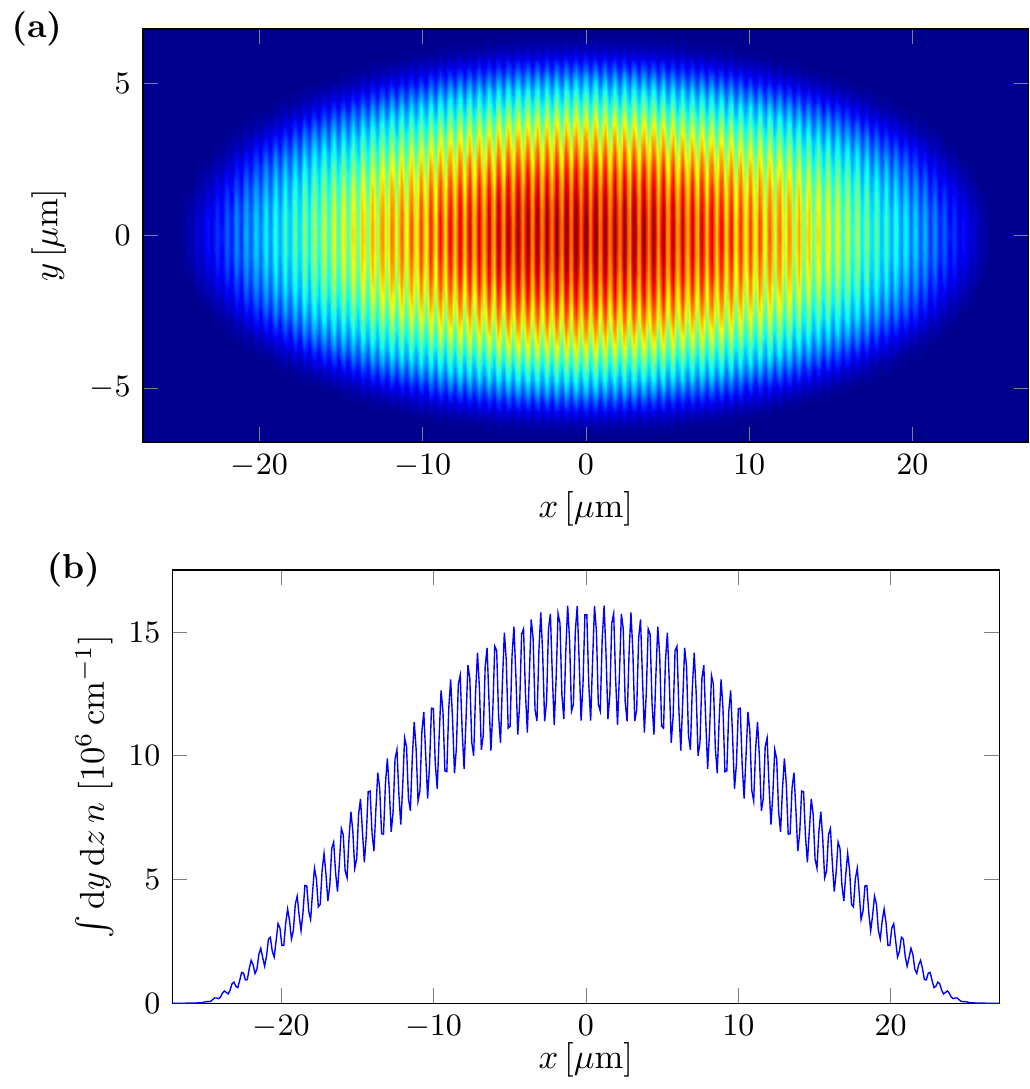}
\caption{Integrated density profiles $\int {\rm d}z \, n$ {\bf (a)}
and $\int {\rm d}y \, {\rm d}z \, n$  {\bf (b)} in the stripe phase,
evaluated in the conditions described in the text, and with traps
separated along $z$ by a distance $d=a_z$, which helps increasing
the visibility of the fringes with respect to
Fig.~\ref{fig:density_prof_d0}.} \label{fig:density_prof_da}
\end{figure}

Quantitative predictions for the novel configuration discussed above
can be obtained by solving numerically the 3D Gross--Pitaevskii
equation. In Figs.~\ref{fig:density_prof_d0} and
\ref{fig:density_prof_da} we show the results for a gas of
$N=4\times 10^4$ $^{87}$Rb atoms confined by an harmonic potential
with frequencies $\left(\omega_x,\omega_y,\omega_z\right) =
2\pi\times\left(25,100,2500\right)\,$Hz, the scattering lengths
equal to those reported in Sec.~\ref{sec:Ground_state}, $k_0=5.54\,
\mu$m$^{-1}$ and $E_r = h\times 1.77\,$kHz consistent with
Ref.~\citen{Lin2011}. Figure~\ref{fig:density_prof_d0} corresponds to
$d=0$, while Fig.~ \ref{fig:density_prof_da} corresponds to
$d=a_z=0.22\, \mu$m, $a_z$ being the harmonic oscillator length
along $z$. In both Figs.~\ref{fig:density_prof_d0} and
\ref{fig:density_prof_da} we have chosen values of the Raman
coupling equal to one half the critical value needed to enter the
plane-wave phase, in order to ensure more stable conditions for the
stripe phase. This corresponds to $\Omega = 0.095\, E_r$ in
Fig.~\ref{fig:density_prof_d0} and to $\Omega = 1.47\, E_r$ in Fig.~
\ref{fig:density_prof_da}. The density plotted in the top panels
corresponds to the 2D density, obtained by integrating the full 3D
density along the $z$ direction; in the bottom panels we show the
double integrated density $\int {\rm d}y \, {\rm d}z \, n$ as a
function of the most relevant $x$ variable. The figures clearly
show that in the conditions of almost equal coupling constants
(Fig.~\ref{fig:density_prof_d0}) the density modulations are very
small, while their effect is strongly amplified in
Fig.~\ref{fig:density_prof_da}, where the interspecies coupling is
reduced with respect to the intraspecies values.

The suggested procedure has also the positive effect of making the
stripe phase more robust against fluctuations of external magnetic
fields. Indeed, the reduction of the interspecies coupling and the
increase of the local density, due to the tight axial confinement,
yield a significant increase of the energy difference between the
stripe and the plane-wave phases. For example, in the case
considered above, for the configuration with a $d=a_z$ displacement
of the two spin layers (Fig.~ \ref{fig:density_prof_da}) a magnetic
detuning of about $0.35 \, E_r$ is required to bring the system into
the spin-polarized phase; in the absence of displacement (Fig.~
\ref{fig:density_prof_d0}) the critical value for the magnetic
detuning is instead much smaller ($\sim 0.001 \, E_r$) \cite{Martone2014b}.

\begin{figure}[t]
\centering
\includegraphics[scale=1]{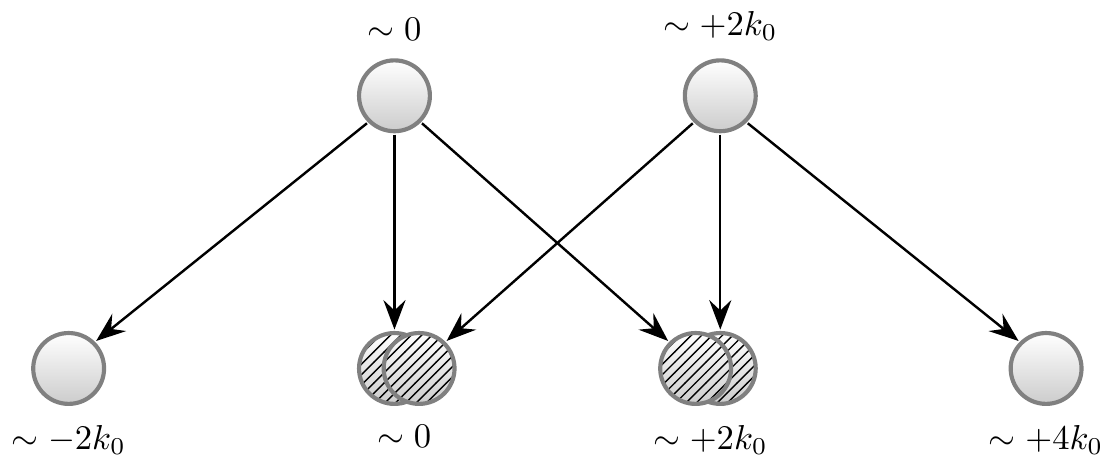}
\caption{Schematical description of the splitting of the spin-down
component of the stripe wave function into different momentum
components caused by a $\pi/2$ Bragg pulse transferring momentum
$2k_1-\epsilon$.} \label{fig:Bragg_scheme}
\end{figure}

\begin{figure}[t]
\centering
\includegraphics[scale=1.0]{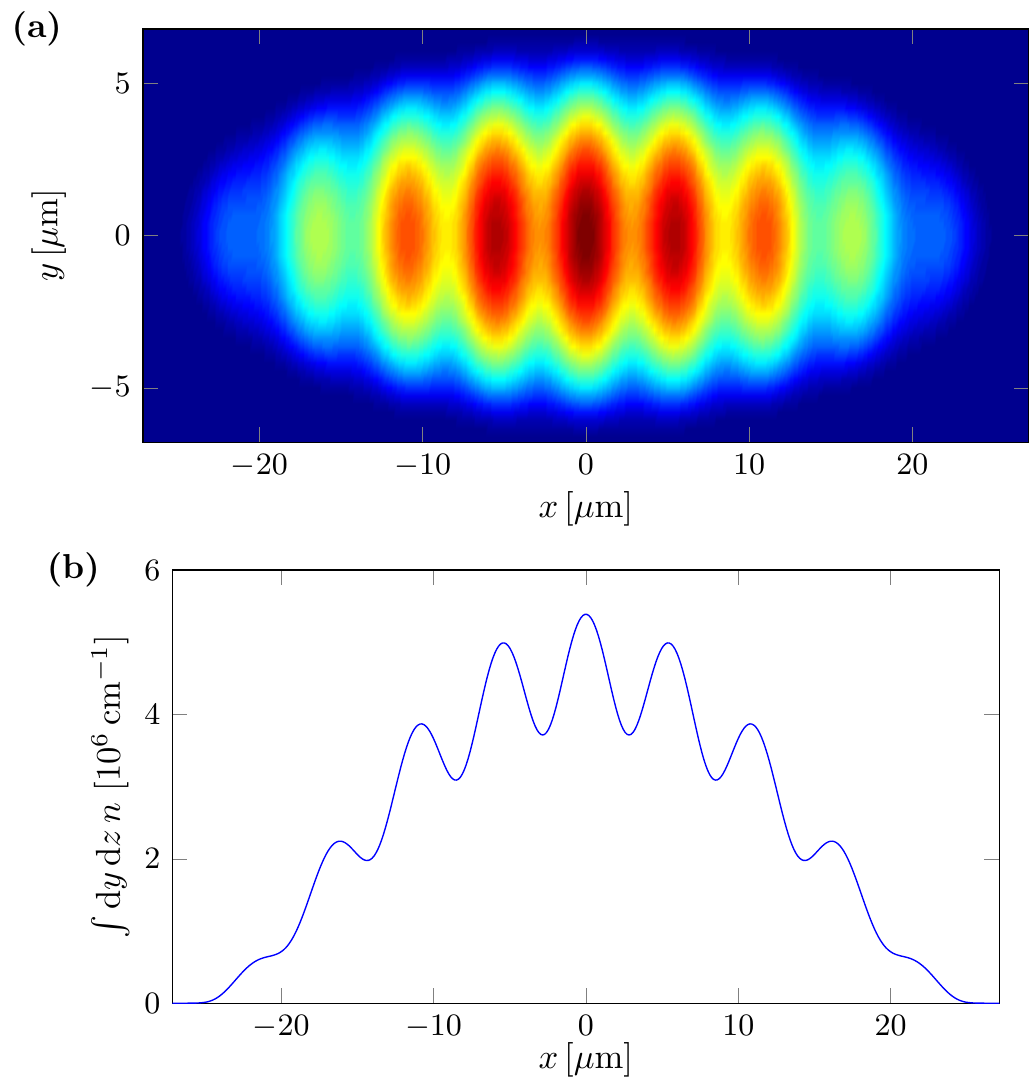}
\caption{Integrated density profiles $\int {\rm d}z \, n$ {\bf (a)}
and $\int {\rm d}y \, {\rm d}z \, n$ {\bf (b)} in the stripe phase,
in the same conditions as Fig.~\ref{fig:density_prof_da}, after the
application of a $\pi/2$ Bragg pulse with transferred momentum
$k_B=1.8\,k_1$.} \label{fig:density_prof_Bragg}
\end{figure}

Let us finally address the problem of the small spatial separation
of the fringes, given by $\pi/k_1$, which turns out to be of the
order of a fraction of a micron in standard conditions. One
possibility to increase the wavelength of the stripes is to lower
the value of $k_0$ by using lasers with a smaller relative incident
angle. In the following we discuss a more drastic procedure which
consists of producing, after the realization of the stripe phase, a
$\pi/2$ Bragg pulse with a short time duration (smaller than the
time $1/E_r$ fixed by the recoil energy), followed by the sudden
release of the trap. This pulse can transfer to the condensate a
momentum $k_B$ or $-k_B$ along the $x$ direction, where $k_B$ is
chosen equal to $2k_1 -\epsilon$ with $\epsilon$ small compared to
$k_1$. The $\pi/2$ pulse has the effect of splitting the condensate
into various pieces, with different momenta. The situation is
schematically shown in Fig.~\ref{fig:Bragg_scheme} for the spin-down
component, where the initial condensate wave function, which in the
stripe phase is a linear combination with canonical momenta $\pm
k_1$, corresponding to momenta $k_0-k_1$ and $k_0+k_1$ in the
laboratory frame, after the Bragg pulse will be decomposed into six
pieces. Two of them, those labeled in the lower part of the figure
with momentum $\sim 0$, will be practically at rest after the pulse
and are able to interfere with fringes of wavelength
$2\pi/\epsilon$, which can easily become large and visible {\it in
situ}. It is worth noticing that these two latter pieces originate
from the two different momentum components of the order parameter
(\ref{eq:ansatz}) in the stripe phase and involve $1/3$ of the total
number of atoms. The corresponding interference effect would be
consequently absent in the plane-wave phase, where only one momentum
component characterizes the order parameter. The other pieces
produced by the Bragg pulse carry much higher momenta and will fly
away rapidly after the release of the trap and of the laser fields.
In Fig.~\ref{fig:density_prof_Bragg} we show a typical behavior of
the density profile obtained by modifying the condensate wave
function in momentum space according to the prescription discussed
above.

\section{Conclusion}
\label{sec:Conclusion}

In this review we have illustrated some relevant static and
dynamic properties of spin-orbit-coupled Bose--Einstein
condensates in the simplest realization of a spin-$1/2$
configuration, characterized by equal Rashba and Dresselhaus
couplings and vanishing or small magnetic detuning. The phase
diagram of these Bose--Einstein condensates is characterized
by the existence of three phases: the stripe,
the plane-wave and the single-minimum phase.
These phases merge in a characteristic tricritical point.
The phase transition between the stripe and the plane-wave
phase has a first-order nature, while the transition between
the plane-wave and the single-minimum phase is of
second order and is characterized by a divergent behavior
of the magnetic polarizability. The stripe phase exhibits
typical density modulations, which are the consequence of
a mechanism of spontaneous breaking of translational
invariance. The three phases discussed in the present
paper exhibit interesting dynamical features, like the
suppression of the dipole oscillation frequency in the presence
of harmonic trapping and of the sound velocity close to the
second-order phase transition, the appearance of a roton minimum
in the plane-wave phase and the occurrence of a double gapless band
structure in the excitation spectrum of the stripe phase. Some of
these features have already been confirmed in recent experiments.
Finally, we have discussed a procedure for the experimental
exploration of the intriguing physics of the stripe phase,
opening new perspectives for the identification of supersolid
phenomena in ultracold atomic gases.

\addcontentsline{toc}{section}{Acknowledgments}
\section*{Acknowledgments}
We wish to thank Lev P. Pitaevskii for many useful and stimulating
discussions. This work has been supported by ERC through the QGBE grant
and by Provincia Autonoma di Trento. The Centre for Quantum Technologies
is a Research Centre of Excellence funded by the Ministry of Education
and National Research Foundation of Singapore.


\appendix

\section{Coefficients in the response function}
\label{app:Response_coeff}
The coefficients in the response function
(\ref{eq:response_uniform}) can be expressed as follows. In phase
II:

\begin{align*}
a &{}=\begin{aligned}[t] &{}- \frac{q^4}{4} + \left[ \left(
k_0^2 + 3 k_1^2 \right) \cos^2\alpha - 2 \left( k_0^2 - G_2 \right)
+ \frac{2 G_2 k_1^2}{k_0^2} \right] q^2 \\
& + 4\left( k_0^2 - 2 G_2 \right) \left[ \left( k_0^2 - k_1^2
\right) \cos^2\alpha - k_0^2 + \frac{2 G_2 k_1^2}{k_0^2}\right] ,
\end{aligned} \\
b_0 &{}=\begin{aligned}[t]  &{}\frac{q^8}{16} -
\left[\left( k_0^2 + k_1^2 \right) \cos^2\alpha - k_0^2 - G_1 +
G_2\right] \frac{q^6}{2}\\
&{}+ \begin{aligned}[t]
\bigg\{ &\! \left(k_0^2 - k_1^2\right)^2 \cos^4 \alpha \\
&{}- 2\left[ k_0^2 \left(k_0^2 - k_1^2\right) + G_1 \left( k_0^2 + 3
k_1^2\right) - G_2 \left(k_0^2 - 5 k_1^2\right) \right] \cos^2\alpha\\
&{}+ k_0^2\left(k_0^2 - 2 G_2\right) + 4 G_1\left(k_0^2 - G_2\right)
+ 2\left(k_0^2 - 2 G_1 - 2 G_2\right) \frac{G_2 k_1^2}{k_0^2} \bigg\}
\, q^4
\end{aligned}\\
&  - 8\left(k_0^2 - 2 G_2\right)
\bigg[ \! \begin{aligned}[t]
&\left(k_0^2 - k_1^2 \right) \left(G_1 + \frac{G_2
k_1^2}{k_0^2}\right) \cos^2\alpha \\
&{}- G_1 k_0^2- \left(k_0^2 - 2 G_1 - 2 G_2\right) \frac{G_2
k_1^2}{k_0^2} \bigg] q^2 ,
\end{aligned}
\end{aligned}\\
b_1 &{}= \begin{aligned}[t] &{}q^4 + 4 \left[\left(k_0^2
- k_1^2\right) \cos^2\alpha + 2\left(G_1 + G_2\right)\right] q^2 +
16 \left(k_0^2 - 2 G_2\right) \left(k_0^2 - k_1^2\right)
\frac{G_2}{k_0^2} ,
\end{aligned}\\
b_2 &{}= \begin{aligned}[t] &{}-\frac{q^4}{2} -
2\left[\left(k_0^2 - 3 k_1^2\right)\cos^2\alpha + k_0^2 + G_1 -
G_2\right] q^2\\
&{} - 4\left(k_0^2 - 2 G_2\right)\left(k_0^2 - \frac{2
G_2 k_1^2}{k_0^2} \right) ,
\end{aligned}
\end{align*}
with $k_1$ given by (\ref{eq:k1_II}). In phase III:
\begin{align*}
a &{}= - \frac{q^4}{4}
- \left(\Omega - k_0^2 \cos^2\alpha + 2 G_2\right)q^2
- \Omega\left[\Omega - 2
\left(k_0^2\cos^2\alpha - 2 G_2\right)\right] , \\
b_0 &{}= \begin{aligned}[t] &{}\;\frac{q^8}{16} +
\left[\Omega - 2\left(k_0^2\cos^2\alpha - G_1 - G_2\right) \right]
\frac{q^6}{4} \\
&{}+ \big[\begin{aligned}[t]
&{}\Omega^2 - 4\left(k_0^2\cos^2\alpha - 2 G_1 - G_2\right) \Omega \\
&{}+ 4\left(k_0^2\cos^2\alpha - 2 G_1\right)
\left(k_0^2 \cos^2 \alpha - 2 G_2\right)\big] \frac{q^4}{4}
\end{aligned}\\
&{}+ 2 G_1 \Omega
\left[\Omega - 2 \left(k_0^2\cos^2\alpha - 2 G_2\right)\right]q^2 ,
\end{aligned}\\
b_1 &{}= 0 , \\
b_2 &{}= -\frac{q^4}{2} - \left[\Omega
+ 2\left(k_0^2\cos^2\alpha + G_1 + G_2\right)\right]q^2 - \Omega
\left(\Omega + 4 G_2\right) .
\end{align*}

\addcontentsline{toc}{section}{References}

\bibliographystyle{ws-rv-van}
\bibliography{Spin_orbit_Review}

\end{document}